\documentclass[%
 reprint,
 amsmath,amssymb,
 aps,
]{revtex4-2}

\usepackage{graphicx}
\usepackage{dcolumn}
\usepackage{bm}
\usepackage{hyperref}
\usepackage{booktabs}
\usepackage{tabularx}

\graphicspath{ {./figures/} }

\usepackage[caption=false]{subfig}
\usepackage{mathtools}
\usepackage{siunitx}
\usepackage{wasysym}
\usepackage{scalerel}

\newcommand{\aap}{Astronomy and Astrophysics}
\newcommand{\icarus}{Icarus}
\newcommand{\mnras}{Monthly Notices of the RAS}
\newcommand{\physrep}{Physics Reports}

\newcommand{\zap}{Zeitschrift fuer Astrophysik}

\include{iss3_commands}

\usepackage{xcolor}

\begin{document}


\title{Timescales of Chaos in the Inner Solar System: \\ 
Lyapunov Spectrum and Quasi-integrals of Motion}

\author{Federico Mogavero}
\thanks{Corresponding author. \href{mailto:federico.mogavero@obspm.fr}{federico.mogavero@obspm.fr}}
\author{Nam H. Hoang}
\thanks{Corresponding author. \href{mailto:nam.hoang-hoai@obspm.fr}{nam.hoang-hoai@obspm.fr}}
\author{Jacques Laskar}
\email{jacques.laskar@obspm.fr}
\affiliation{IMCCE, CNRS UMR 8028, Observatoire de Paris, Université PSL, Sorbonne Université \\
77 Avenue Denfert-Rochereau, 75014 Paris, France}

\date{\today}

\begin{abstract}
Numerical integrations of the Solar System reveal a remarkable stability of the orbits of the inner planets over billions of years, in spite of 
their chaotic variations characterized by a Lyapunov time of only 5 million years and the lack of integrals of motion able to constrain their dynamics. 
To open a window on such long-term behavior, we compute the entire Lyapunov spectrum of a forced secular model of the inner planets. We uncover 
a hierarchy of characteristic exponents that spans two orders of magnitude, manifesting a slow-fast dynamics with a broad separation of timescales. 
A systematic analysis of the Fourier harmonics of the Hamiltonian, based on computer algebra, reveals three symmetries that characterize 
the strongest resonances responsible for the orbital chaos. These symmetries are broken only by weak resonances, 
leading to the existence of quasi-integrals of motion that are shown to relate to the smallest Lyapunov exponents. 
A principal component analysis of the orbital solutions independently confirms that the quasi-integrals are among the slowest degrees of freedom of the dynamics. 
Strong evidence emerges that they effectively constrain the chaotic diffusion of the orbits, playing a crucial role in the statistical stability 
over the Solar System lifetime. 
\end{abstract}

\maketitle


\section{Introduction}
\label{sec:intro}
The planetary orbits in the inner Solar System (ISS) are chaotic, with a Lyapunov time distributed around 5 million years 
(Myr) \citep{Laskar1989,Laskar1990,Sussman1992,Mogavero2021}. Still, they are statistically very stable over a timescale 
that is a thousand times longer. The probability that the eccentricity of Mercury exceeds 0.7, leading to catastrophic events 
(i.e., close encounters, collisions, or ejections of planets), 
is only about 1\% over the next 5 billion years (Gyr) \citep{Laskar2008,Laskar2009,Hoang2022}. The dynamical half-life of 
Mercury orbit has recently been estimated at 30--40 billion years \citep{Mogavero2021,Hoang2022}. A disparity of nearly 
four orders of magnitude between the Lyapunov time and the timescale of dynamical instability is intriguing, since the 
chaotic variations of the orbits of the inner planets cannot be constrained \textit{a priori}. 
While the total energy and angular momentum of the Solar System are conserved, the disproportion of masses between the outer 
and inner planets implies that unstable states of the ISS are in principle easily realizable through exchanges of these quantities. 
The surprising stability of the ISS deserves a global picture in which it can emerge more naturally. 

To our knowledge, the only study addressing the timescale separation in the long-term dynamics of the ISS is based 
on the simplified secular dynamics of a massless Mercury \citep{Batygin2015}: All the other planets are frozen on regular 
quasi-periodic orbits; secular interactions are expanded to first order in masses and degree 4 in eccentricities 
and inclinations; an \textit{a priori} choice of the relevant terms of the Hamiltonian is made.   
The typical instability time of about 1 Gyr \citep{Batygin2015,Woillez2020} is, however, too short and in significant contrast 
with realistic numerical integrations of the Solar System, which show a general increase of the instability rate with the 
complexity of the dynamical model \citep{Hoang2022}. 
We have shown that truncating the secular Hamiltonian of the ISS at degree 4 in eccentricities and inclinations 
results in an even more stable dynamics, with an instability rate at 5 Gyr that drops by orders of magnitude when compared 
to the full system\footnote{The dynamics truncated at degree 4 produces nevertheless the same chaos of the full system, 
as measured by the finite-time maximum Lyapunov exponent \citep{Mogavero2022}.}. 
From the perspective of these latest findings, the small probability of 1\% of an instability over the age of the Solar System 
may be naturally regarded as a perturbative effect of terms of degree 6 and higher. Clearly, the striking stability 
of the dynamics at degree 4 is even more impressive in the present context, and remains to be explained. 

A strong separation in dynamical timescales is not uncommon among classical quasi-integrable systems \citep[e.g.,][]{Milani1992,Morbidelli1996}. 
This is notably evinced by the Fermi-Pasta-Ulam-Tsingou (FPUT) problem, which deals with a chain of coupled weakly-anharmonic oscillators \citep{Fermi1955}. 
Far from Kolmogorov-Arnold-Moser (KAM) and Nekhoroshev regimes (as is likely to be pertinent to the ISS, see Sect.~\ref{sec:spectrum}), 
one can generally state that the exponential divergence of close trajectories occurring over a Lyapunov time 
is mostly tangent to the invariant tori defined by the action variables 
of the underlying integrable problem, and hence contributes little to the diffusion in the action space \citep{Lam2014,Goldfriend2019}. 
In other words, the Lyapunov time and the diffusion/instability time scale differently with the size of the terms 
that break integrability, and this can result in very different timescales \citep{Morbidelli1996}. 
However, this argument is as general as poorly 
satisfactory in addressing quantitatively the timescale separation in a complex problem as the present one. 
Moreover, even though order-of-magnitude estimates of the chaotic diffusion in the ISS suggest that it may take 
hundreds of million years to reach the destabilizing secular resonance $g_1-g_5$\footnote{This resonance involves 
the fundamental precession frequencies of Mercury and Jupiter perihelia \citep{Laskar2008,Batygin2008,Boue2012}.}, 
the low probability of an instability over 5 Gyr still remains unexplained \citep{Mogavero2021}. 
Establishing more precisely why the ISS is statistically stable over a timescale comparable to its age 
is a valuable step in understanding the secular evolution of planetary systems through metastable states 
\citep{Laskar1996,Mogavero2021}\footnote{``At each stage of its evolution, the system should have a time of stability
comparable with its age'' \citep{Laskar1996}}. 
With its 8 secular degrees of freedom (d.o.f.), this system also constitutes a peculiar bridge between the low-dimensional dynamics 
often addressed in celestial mechanics and the systems with a large number of bodies studied in statistical mechanics: 
It cannot benefit from the straightforward application of standard methods of the two fields \citep[e.g.,][Appendix~A]{Mogavero2022}. 

This work aims to open a window on the long-term statistical behavior of the inner planet orbits. 
Section~\ref{sec:model} briefly recalls the dynamical model of forced secular ISS introduced in Ref.~\citep{Mogavero2021}. 
Section~\ref{sec:spectrum} presents the numerical computation of its Lyapunov spectrum. Section~\ref{sec:quasi_integrals} 
introduces the quasi-symmetries of the resonant harmonics of the Hamiltonian and the corresponding quasi-integrals (QIs) of motion. 
Section~\ref{sec:PCA} establishes a geometric connection between the quasi-integrals and the slowest d.o.f. of the dynamics via a principal 
component analysis (PCA) of the orbital solutions. Section~\ref{sec:implications} states the implications of the new findings on the 
long-term stability of the ISS. We finally discuss the connections with other classical quasi-integrable systems and the methods 
used in this work. 

\begin{table*}[ht]
\caption{\label{tab:table_Hamiltonians}
Summary of the different models of forced secular ISS considered in this work. 
Gauss's dynamics results from first-order averaging of the $N$-body Hamiltonian over the mean longitudes of the planets. 
The dynamics generated by $\Hiss_{2n}$ and $\Huv_{2n}$ are practically equivalent and treated as such. 
The $\simpleH{2n}$ models are introduced and discussed in Sec.~\ref{sec:new_truncation}.}
\begin{tabularx}{\linewidth}{l X r}
\hline 
\hline 
Hamiltonian & \multicolumn{1}{c}{Description}  & Reference \\
\hline 
$\Hiss$ & Gauss's dynamics in complex Poincaré variables & Eq.~\eqref{eq:ham_sec_inn}  \\
$\Hiss_{2n}$ & Truncation of Gauss's dynamics at total degree $2n$ in eccentricities and inclinations & Ref.~\cite{Mogavero2021}  \\
$\Huv_{2n}$ & Truncated dynamics in the action-angle variables of the Laplace-Lagrange dynamics $\Hiss_2$ & Eq.~\eqref{eq:hamiltonian} and Ref.~\cite{Mogavero2021} \\
$\simpleH{2n}$ & Fourier harmonics that involve outer planet modes other than $g_5$ are dropped from $\Huv_{2n}$ & Eq.~\eqref{eq:simpleH} \\
\hline 
\end{tabularx}
\end{table*}

\section{Dynamical model}
\label{sec:model}
The long-term dynamics of the Solar System planets consists essentially of the slow precession of their perihelia and nodes, 
driven by secular, orbit-averaged gravitational interactions \cite{Laskar1990,Laskar2004}. At first order in planetary masses, 
the secular Hamiltonian corrected for the leading contribution of general relativity reads \citep[e.g.][]{Morbidelli2002,Mogavero2021} 
\begin{equation}
\label{eq:ham_sec}
\Hsec = - \sum_{i=1}^8  
\left[ \sum_{l=1}^{i-1} \left< \frac{G m_i m_l}{\| \vec{r}_i - \vec{r}_l \|} \right> 
+ \frac{3 G^2 m_0^2 m_i}{c^2 a_i^2 \sqrt{1-e_i^2}} \right] .
\end{equation}
The planets are indexed in order of increasing semi-major axes $(a_i)_{i=1}^8$, $m_0$ and $m_i$ are the Sun and
planet masses, respectively, $e_i$ the eccentricities, $G$ the gravitational constant and $c$ the speed of light. 
The vectors $\vec{r}_i$ are the heliocentric positions of the planets, and the bracket operator represents the averaging 
over the mean longitudes resulting from the elimination of the non-resonant Fourier harmonics 
of the $N$-body Hamiltonian \citep{Morbidelli2002,Mogavero2021}. Hamiltonian~\eqref{eq:ham_sec} generates Gauss's 
dynamics of Keplerian rings \citep{Gauss1818,Mogavero2021}, whose semi-major axes $a_i$ are constants of motion of the 
secular dynamics. 

By developing the 2-body perturbing function \citep{Laskar1991,Laskar1995} in the computer algebra system \TRIP{} \citep{Gastineau2011,TRIP}, 
the secular Hamiltonian can be systematically expanded in series of the Poincaré rectangular coordinates in complex~form, 
\begin{equation}
\label{eq:poincare_vars}
\begin{aligned}
x_i &= \sqrt{\Lambda_i} \sqrt{ 1 - \sqrt{1-e_i^2} } \E^{j \varpi_i}, \\
y_i &= \sqrt{2 \Lambda_i} \left(1- e_i^2\right)^{\frac{1}{4}} \sin(\inc_i/2) \E^{j \Omega_i} ,
\end{aligned}
\end{equation}
where $\Lambda_i = \mu_i [G(m_0 + m_i) a_i]^{1/2}$, $\mu_i = m_0 m_i / (m_0 + m_i)$ being the reduced masses of the planets, 
$\inc_i$ the inclinations, $\varpi_i$ the longitudes of the perihelia and $\Omega_i$ the longitudes of the nodes\footnote{$\E$ 
represents the exponential operator and $j$ stands for the imaginary unit. The overline on variables denotes complex conjugate.}. 
Pairs $(x_i,-j \overline{x}_i)$ and $(y_i,-j \overline{y}_i)$ are canonically conjugate momentum-coordinate variables. 
When truncating at a given total degree $2n$ in eccentricities and inclinations, the expansion provides 
Hamiltonians $\Hsec_{2n} = \Hsec_{2n}[(x_i, \bar{x}_i, y_i, \bar{y}_i)_{i=1}^8]$ that are multivariate polynomials. 

Valuable insight into the dynamics of the inner planets is provided by the model of a forced ISS recently proposed \citep{Mogavero2021}. 
It exploits the great regularity of the long-term motion of the outer planets \citep{Laskar1990,Laskar2004,Hoang2021} to 
predetermine their orbits to a quasi-periodic form: 
\begin{eqnarray}
\label{eq:qp_decomposition}
\hspace{-5mm} 
x_i(t) = \sum_{l=1}^{M_i} \tilde{x}_{il} \, 
\E^{j \vec{m}_{il} \cdot \out{\vec{\omega}} t}, \;
y_i(t) = \sum_{l=1}^{N_i} \tilde{y}_{il} \, 
\E^{j \vec{n}_{il} \cdot \out{\vec{\omega}} t} ,
\end{eqnarray}
for $i \in \{5,6,7,8\}$, where $t$ denotes time, $\tilde{x}_{il}$ and $\tilde{y}_{il}$ are complex amplitudes, 
$\vec{m}_{il}$ and $\vec{n}_{il}$ integer vectors, and $\out{\vec{\omega}} = (g_5,g_6,g_7,g_8,s_6,s_7,s_8)$ 
represents the septuple of the constant fundamental frequencies of the outer orbits. 
Frequencies and amplitudes of this Fourier decomposition are established numerically by frequency analysis \citep{Laskar1988,Laskar2005} 
of a comprehensive orbital solution of the Solar System \citep[Appendix~D]{Mogavero2021}. 
Gauss's dynamics of the forced ISS is obtained by substituting the predetermined time dependence in Eq.~\eqref{eq:ham_sec}, 
\begin{equation}
\label{eq:ham_sec_inn}
\Hiss = 
\Hsec[(x_i,y_i)_{i=1}^4, (x_i = x_i(t), y_i = y_i(t))_{i=5}^8] ,
\end{equation}
so that $\Hiss = \Hiss[(x_i,y_i)_{i=1}^4,t]$. The resulting dynamics consists of two d.o.f. for each inner planet, 
corresponding to the $x_i$ and $y_i$ variables, respectively. Therefore, the forced secular ISS is described by 8 d.o.f. and an 
explicit time dependence. As a result of the forcing from the outer planets, no trivial integrals of motion exist and its orbital 
solutions live in a 16-dimensional phase space. 

A truncated Hamiltonian $\Hiss_{2n}$ for the forced ISS is readily obtained by substituting Eq.~\eqref{eq:qp_decomposition} in 
the truncated Hamiltonian $\Hsec_{2n}$ of the entire Solar System. At the lowest degree, $\Hiss_2$ generates a linear, forced
Laplace-Lagrange (LL) dynamics. This can be analytically integrated by introducing complex proper mode variables 
$(u_i, v_i)_{i=1}^4$ via a time-dependent canonical transformation $(x_i,-j \overline{x}_i) \rightarrow (u_i,-j \overline{u}_i)$, 
$(y_i,-j \overline{y}_i) \rightarrow (v_i,-j \overline{v}_i)$ \citep{Mogavero2021}. 
Action-angle pairs $(\Chi_i,\chi_i)$, $(\Psi_i,\psi_i)$ are introduced as 
\begin{equation}
\label{eq:AAvars}
u_i = \sqrt{\Chi_i} \E^{-j \chi_i}, \;
v_i = \sqrt{\Psi_i} \E^{-j \psi_i} . 
\end{equation}
When expressed in the proper modes, the truncated Hamiltonian can be expanded as a finite Fourier series: 
\begin{equation}
\label{eq:hamiltonian}
\Huv_{2n}(\vecI,\vectheta,t) = 
\sum_{\vec{k},\vecl}
\fourcoeff{2n}{k}{\ell} (\vecI)
\E^{j \left( \vec{k} \cdot \vec{\theta} + \vec{\ell} \cdot \vec{\phi}(t) \right)} ,
\end{equation}
where $\vecI = (\vecChi,\vecPsi)$ and $\vectheta = (\vecchi,\vecpsi)$ are the 8-dimensional vectors of 
the action and angle variables, respectively, and we introduce the external angles $\vec{\phi}(t) = -\out{\vec{\omega}} t$. 
The wave vectors $(\veck, \vecl)$ belong to a finite subset of $\mathbb{Z}^8 \times \mathbb{Z}^7$. 
At degree 2, one has $\Huv_2 = -\LL{\vec{\omega}} \cdot \vecI$, 
where $\LL{\vec{\omega}} = (\LL{\vec{g}}, \LL{\vec{s}}) \in \mathbb{R}^4 \times \mathbb{R}^4$ are the LL fundamental 
precession frequencies of the inner planet perihelia and nodes. Hamiltonian $\Huv_{2n}$ is in quasi-integrable form. 

The quasi-periodic form of the outer orbits in Eq.~\eqref{eq:qp_decomposition} contains harmonics of order higher 
than one, that is, $\| \vec{m}_{il} \|_1 > 1$ and $\| \vec{n}_{il} \|_1 > 1$ for some $i$ and $l$, where $\| \cdot \|_1$ 
denotes the 1-norm. Therefore, the dynamics of $\Hiss_{2n}$ and $\Huv_{2n}$ are not exactly the same \citep{Mogavero2021}. 
Still, the difference is irrelevant for the results of this work, so we treat the two Hamiltonians as equivalent 
from now on. 
Despite the simplifications behind Eqs.~\eqref{eq:ham_sec} and \eqref{eq:qp_decomposition}, the forced secular ISS has been 
shown to constitute a realistic model that is consistent with the predictions of reference integrations of the 
Solar System \citep{Laskar1990,Laskar2004,Laskar2008,Laskar2009}. It correctly reproduces the finite-time maximum Lyapunov 
exponent (FT-MLE) and the statistics of the high eccentricities of Mercury over 5 Gyr \citep{Mogavero2021}. 
Table~\ref{tab:table_Hamiltonians} presents a summary of the different Hamiltonians and corresponding dynamics 
we consider in this work. 

\section{Lyapunov spectrum}
\label{sec:spectrum}
Ergodic theory provides a way, through the Lyapunov characteristic exponents (LCEs), to introduce a fundamental set of timescales 
for any differentiable dynamical system $\dot{\vecz} = \vecF(\vecz, t)$ defined on a phase space $\mathcal{P} \subseteq \mathbb{R}^P$ 
\citep{Oseledec1968,Eckmann1985,Gaspard1998,Skokos2010}. 
If $\vec{\Phi}(\vecz,t)$ denotes the associated flow and $\vecz(t) = \vec{\Phi}(\vecz_0,t)$ the orbit that emanates from 
the initial condition $\vecz_0$, the LCEs $\lambda_1 \geq \lambda_2 \geq \dots \geq \lambda_P$ are the logarithms of the eigenvalues 
of the matrix $\matLambda(\vecz_0)$ defined as 
\begin{equation}
\label{eq:matLambda}
\lim_{t \rightarrow \infty} \left( \matM(\vecz_0,t)^\mathrm{T} \matM(\vecz_0,t) \right)^{1/2t} = 
\matLambda(\vecz_0) ,
\end{equation}
where $\matM(\vecz_0,t) = \partial \vec{\Phi} / \partial \vecz_0$ is the fundamental matrix and $\mathrm{T}$ stands for transposition \citep{Eckmann1985,Gaspard1998}. 
Introducing the Jacobian $\matJ = \partial \vecF / \partial \vecz$, the fundamental matrix allows us to write the solution of the 
variational equations $\dot{\veczeta} = \matJ(\vecz(t),t) \veczeta$ as $\veczeta(t) = \matM(\vecz_0,t) \veczeta_0$, 
where $\veczeta(t) \in \mathcal{T}_{\vecz(t)}\mathcal{P}$ belongs to the tangent space of $\mathcal{P}$ at point $\vecz(t)$ 
and $\veczeta_0 = \veczeta(0)$. The multiplicative ergodic theorem of \citet{Oseledec1968} states that if $\rho$ is an ergodic 
(i.e. invariant and indecomposable) measure for the time evolution and has compact support, then the limit in Eq.~\eqref{eq:matLambda} 
exists for $\rho$-almost all $\vecz_0$, and the LCEs are $\rho$-almost everywhere constant and only depend on $\rho$ \citep{Eckmann1985}. 
Moreover, one has 
\begin{equation}
\label{eq:LCEs}
\lim_{t \rightarrow \infty} \frac{1}{t} \log \Vert \matM(\vecz_0,t) \veczeta_0 \Vert = \lambda^{(i)} \enspace
\textrm{if} \enspace \veczeta_0 \in E_{\vecz_0}^{(i)} \setminus E_{\vecz_0}^{(i+1)} , 
\end{equation}
for $\rho$-almost all $\vecz_0$, where $\lambda^{(1)} > \lambda^{(2)} > \dots$ are the LCEs without repetition by multiplicity, 
and $E_{\vecz_0}^{(i)}$ is the subspace of $\mathbb{R}^P$ corresponding to the eigenvalues of $\matLambda(\vecz_0)$ that are smaller than 
or equal to $\exp\lambda^{(i)}$, with $\mathcal{T}_{\vecz_0}\mathcal{P} = E_{\vecz_0}^{(1)} \supset E_{\vecz_0}^{(2)} \supset \cdots$. 
The specific choice of the $\mathbb{R}^P$-norm $\Vert \cdot \Vert$ in Eq.~\eqref{eq:LCEs} is irrelevant 
\citep{Eckmann1985,Skokos2010}. 
Once the LCEs have been introduced, a characteristic timescale can be defined from each positive exponent as $\lambda_i^{-1}$. 
In the case of the maximum Lyapunov exponent $\lambda_1$, the corresponding timescale is commonly called the Lyapunov~time. 

For a Hamiltonian system with $p$ d.o.f. (i.e., $P=2p$), the fundamental matrix is symplectic and the set of LCEs 
is symmetric with respect to zero, that is, 
\begin{equation}
\label{eq:LCE_symmetry}
\Delta\lambda_i \vcentcolon = \lambda_i + \lambda_{2p-i+1} = 0 \enspace \textrm{for all} \enspace 1 \leq i \leq p .
\end{equation}
If the Hamiltonian is time independent, a pair of exponents vanishes. In general, the existence of an integral of motion 
$\mathfrak{C} = \mathfrak{C}(\vec{z})$ implies a pair of null exponents, one of them being associated with the direction 
of the tangent space that is normal to the surface of constant $\mathfrak{C}$ \citep[e.g.][]{Gaspard1998}. 

The ISS is a clear example of a dynamical system that is out of equilibrium. Its phase-space density diffuses seamlessly over 
any meaningful timescale \citep{Laskar2008,Hoang2021}. Therefore, the infinite time limit in Eq.~\eqref{eq:matLambda} is not 
physically relevant. The non-null probability of a collisional evolution of the inner planets \citep{Laskar1994,Laskar2008,Batygin2008,Laskar2009} 
implies that such limit does not even exist as a general rule. Most of the orbital solutions stemming from the current knowledge of 
the Solar System are indeed asymptotically unstable \citep{Mogavero2021,Hoang2022}. Physically relevant quantities are the 
finite-time LCEs (FT-LCEs), $\lambda_i(\vecz_0,t)$, defined from the eigenvalues $\mathfrak{m}_1 \geq \mathfrak{m}_2 \geq \dots \geq \mathfrak{m}_P$ 
of the time-dependent symmetric positive-defined matrix $\matM(\vecz_0,t)^\mathrm{T} \matM(\vecz_0,t)$ as 
\begin{equation}
\label{eq:FT_LCEs}
\lambda_i(\vecz_0,t) = \frac{1}{2t} \log \mathfrak{m}_i(\vecz_0,t) .
\end{equation}
The time dependence of the phase-space density translates in the fact that no ergodic measure is realized by the dynamics, and 
the FT-LCEs depend on the initial condition $\vecz_0$ in a non-trivial way\footnote{The FT-LCEs do not depend on any tangent 
vector $\veczeta_0$.}. 

\begin{figure*}
\subfloat[\label{fig:LCE4_1} $\Hiss_4$]{\includegraphics[width=0.99\columnwidth]{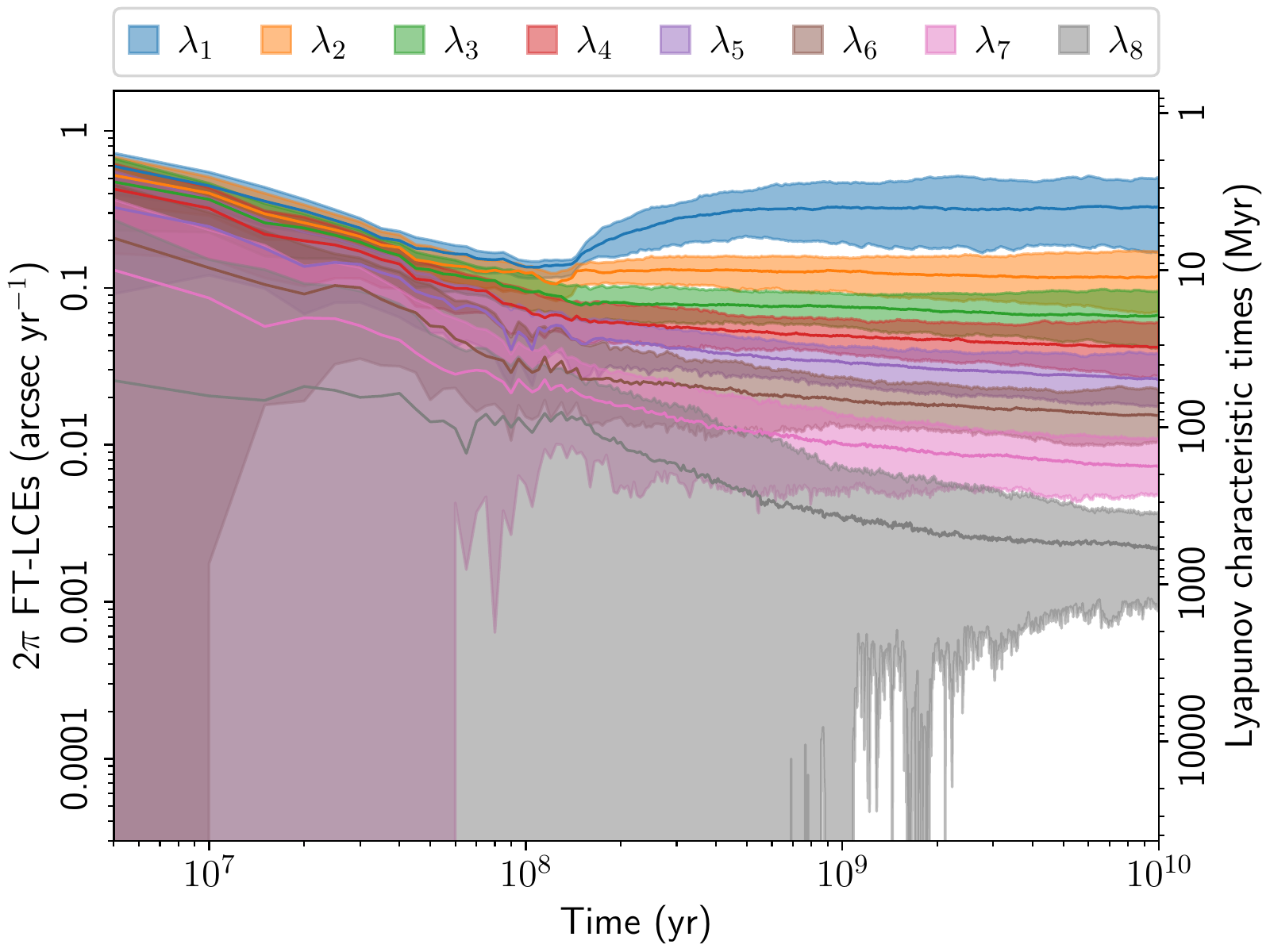}} 
\hphantom{ciao}
\subfloat[\label{fig:LCE4_2} $\simpleH{4}$]{\includegraphics[width=0.99\columnwidth]{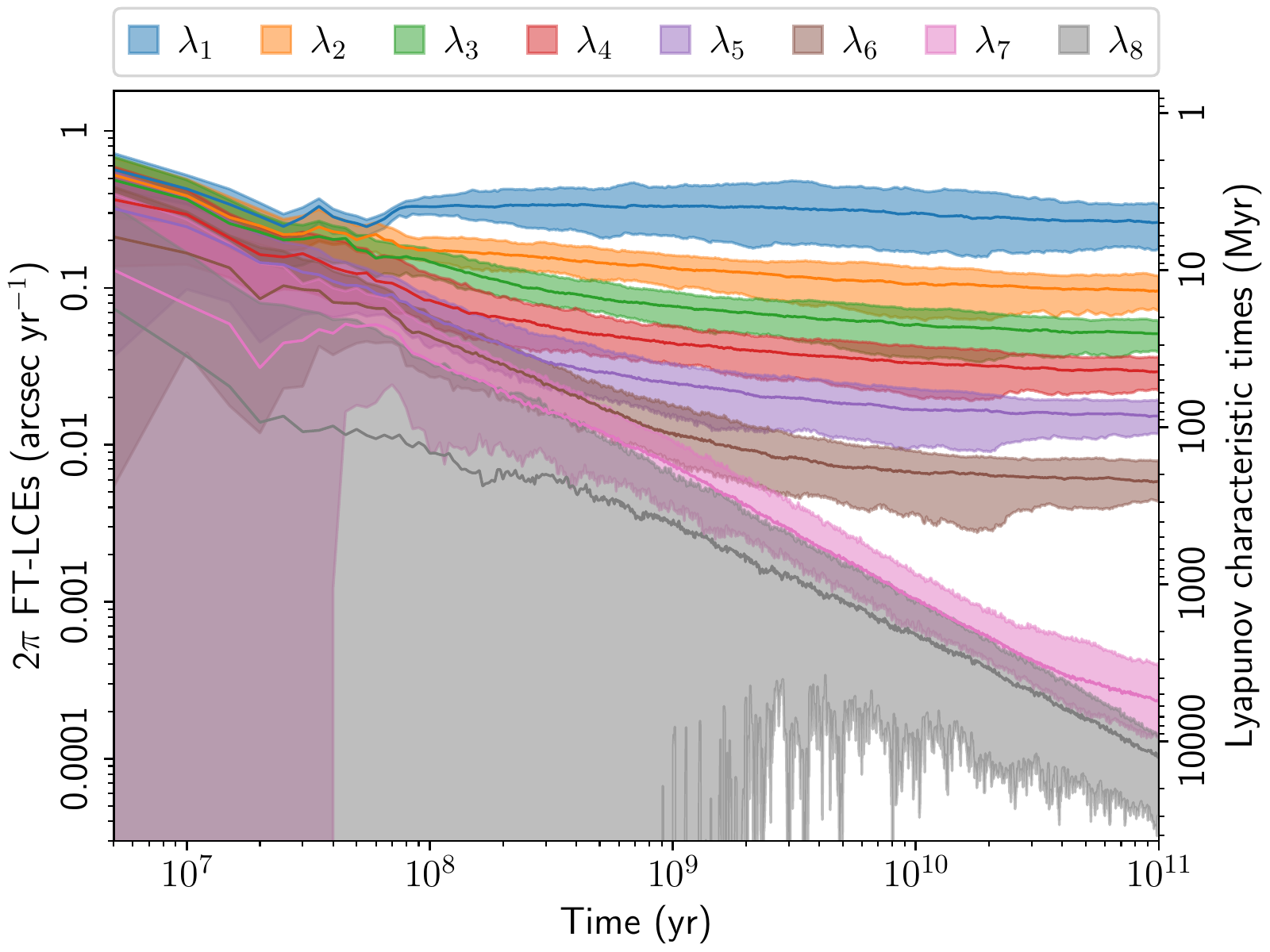}}
\caption{Positive FT-LCEs $\lambda_i$ of the forced secular ISS from Hamiltonians $\Hiss_4$ (a) 
and $\simpleH{4}$ [Eq.~\eqref{eq:simpleH}] (b), and corresponding characteristic timescales $\lambda_i^{-1}$. 
The bands represent the [5th, 95th] percentile range of the marginal PDFs estimated 
from an ensemble of 150 stable orbital solutions with very close initial conditions. 
The lines denote the distribution medians. The $\simpleH{4}$ model is introduced and discussed in Sec.~\ref{sec:new_truncation}.}
\label{fig:LCE4}
\end{figure*}

The FT-MLE of the forced secular ISS has been numerically computed over 5 Gyr for an ensemble of stable orbital 
solutions of the Hamiltonian $\Hiss$ with initial conditions very close to their nominal values\footnote{All the 
initial conditions mentioned in this work are sampled from a multidimensional Gaussian distribution that is centered 
at the nominal initial conditions of Gauss's dynamics \citep[Appendix~D]{Mogavero2021} and has a relative width 
of $10^{-9}$ (see \citep[Appendix~C]{Mogavero2022} and \citep[Sect.~3]{Hoang2022} for details).}. 
Its long-term distribution is quite large and does not shrink over time \citep[Fig.~3]{Mogavero2021}. At 5 Gyr, the probability density function (PDF) 
of the Lyapunov time peaks at around 4 Myr, it decays very fast below 2 Myr, while its 99th percentile reaches 10 Myr 
\citep[Fig.~4]{Mogavero2021}. 
The significant width of the distribution relates to the aforementioned out-of-equilibrium dynamics of the ISS, as the FT-MLE of 
each orbital solution continues to vary over time. The dependence of the exponent on the initial condition is associated with 
the non-ergodic exploration of the phase space by the dynamics. 
As a remark, the fact that the lower tail of the FT-MLE distribution, estimated from more than 1000 solutions, 
does not extend to zero implies that invariant (KAM) tori are rare in a neighborhood of the nominal initial conditions 
(if they exist at all). 
This fact excludes that the dynamics is in a Nekhoroshev regime \citep{Morbidelli1995,Morbidelli1996}, in agreement with the indications 
of a multidimensional resonance overlapping at the origin of chaos \citep{Laskar1992,Mogavero2022}. In such a case, 
the long dynamical half-life of the ISS should not be interpreted in terms of an exponentially-slow Arnold diffusion. 

Computations of the FT-MLE of the Solar System planets have been reported for more than thirty years \citep{Laskar1989,Sussman1992}. 
However, the retrieval of the entire spectrum of exponents still represents a challenging task. Integrating an $N$-body orbital 
solution for the Sun and the eight planets that spans 5 Gyr requires the order of a month of wall-clock time 
\citep[e.g.][]{Brown2020}. 
The computation by a standard method of the entire Lyapunov spectrum for a system with $p$ d.o.f. also requires the simultaneous 
time evolution of a set of $2p$ tangent vectors \citep[][]{Benettin1978}. On the top of that, a computation of the exponents for an 
ensemble of trajectories is advisable for a non-ergodic dynamics \citep{Mogavero2021}. These considerations show how demanding the computation 
of the Lyapunov spectrum of the Solar System planets is. 
By contrast, a 5-Gyr integration of the forced ISS takes only a couple of hours for Gauss's dynamics ($\Hiss$) and a few minutes at 
degree 4 ($\Hiss_4$). This dynamical model thus provides a unique opportunity to compute all the FT-LCEs that are mainly related to 
the secular evolution of the inner orbits. 

We compute the Lyapunov spectrum of the truncated forced ISS using the standard method of \citet{Benettin1980}, 
based on Gram-Schmidt orthogonalization. Manipulation of the truncated Hamiltonian $\Hiss_{2n}$ in \TRIP{} allows us to systematically 
derive the equations of motion and the corresponding variational equations, which we integrate through an Adams PECE method 
of order 12 and a  timestep of 250 years. 
Parallelization of the time evolution of the 16 tangent vectors, between two consecutive reorthonormalization steps of the 
\citet{Benettin1980} algorithm, significantly reduces the computation time. 
Figure~\ref{fig:LCE4_1} shows the positive FT-LCEs expressed as angular frequencies over the next 10 Gyr for the Hamiltonian 
truncated at degree 4. The FT-LCEs are computed for 150 stable solutions, with initial conditions very close to the nominal values of Gauss's dynamics 
and random sets of initial tangent vectors \citep[][Appendix~C]{Mogavero2022}. The figure shows the [5th, 95th] percentile range of the marginal PDF 
of each exponent estimated from the ensemble of solutions. 
For large times, the exponents of each solution become independent of the initial tangent vectors, the renormalization time, and the norm chosen 
for the phase-space vectors (see Appendix~\ref{sec:LCE_convergence} and Fig.~\ref{fig:LCE4_test1}). In this asymptotic regime, the \citet{Benettin1980} algorithm purely retrieves 
the FT-LCEs as defined in Eq.~\eqref{eq:FT_LCEs}, and the width of their distributions only reflects the out-of-equilibrium dynamics of the system. The convergence 
of our numerical computation is also assessed by verifying the symmetry of the spectrum stated in Eq.~\eqref{eq:LCE_symmetry} 
(see Appendix~\ref{sec:LCE_convergence} and Fig.~\ref{fig:LCE4_test2}). 

The spectrum in Fig.~\ref{fig:LCE4_1} has distinctive features. A set of intermediate exponents follow the MLE, ranging from 0.1 to 0.01\arcsecyr, 
while the smallest ones fall below 0.01\arcsecyr. Figure~\ref{fig:LCE4_1} reveals the existence of a hierarchy of exponents and 
corresponding timescales that spans two orders of magnitude, down to a median value of $\lambda_8^{-1} \approx 500$ Myr. 
The number of positive exponents confirms that no integral of motion exists, as one may expect from the forcing of the outer planets. 
We also compute the spectrum for the Hamiltonian truncated at degree 6. As shown in Appendix~\ref{sec:LCE_6} (Fig.~\ref{fig:LCE6}), 
the asymptotic distributions of the exponents are very similar to those at degree 4. 
This result suggests that long-term diffusion of the phase-space density is very close in the two cases. 
The different instability rates of the two truncated dynamics mainly relates to the geometry of the instability boundary, which is closer 
to the initial position of the system for $\Hiss_6$ than for $\Hiss_4$ \citep{Hoang2022}. 

The relevance of the Lyapunov spectrum in Fig.~\ref{fig:LCE4_1} emerges from the fact that the existence of an integral of motion 
implies a pair of vanishing exponents. This is a pivotal point: By a continuity argument, the presence of positive exponents 
much smaller than the leading one constitutes a compelling indication that there are dynamical quantities 
whose chaotic decoherence over initially very close trajectories takes place over timescales much longer than the Lyapunov time. 
In the long term, such quantities should diffuse much more slowly than any LL action variable. Therefore, Fig.~\ref{fig:LCE4_1} 
suggests that the secular orbits of the inner planets are characterized by a slow-fast dynamics that is much more pronounced 
than the well-known timescale separation arising from the LL integrable approximation. The existence of slow quantities, which are 
\textit{a priori} complicated functions of the phase-space variables, is crucial in the context of finite-time stability, 
as they can effectively constrain the long-term diffusion of the phase-space density toward the unstable states. 
The next section addresses the emergence of these slow quantities from the symmetries of the Fourier harmonics that compose the Hamiltonian. 

\begin{table}
\caption{\label{tab:table1}
Top of ranking $\R_1$. First 30 resonant harmonics of $\Huv_{10}$ along the 5-Gyr nominal solution of Gauss's dynamics, in order 
of decreasing time median of the resonance half-width $\halfwidth$ (\arcsecyrtext{}). Adapted from Table~2 of Ref.~\citep{Mogavero2022}.} 
\begin{ruledtabular}
\begin{tabular}{rrrrc}
$i$ & Fourier harmonic $\F_i$ & $\O_i$ & $\tau_i^\mathrm{res}$ & $\halfwidth_i$ \\
\colrule
\rule{0pt}{2.3ex}
\input{tables/strongest_resonances.out}
\end{tabular}
\end{ruledtabular}
\footnotetext[1]{$\O$ is the order of the harmonic.}
\footnotetext[2]{$\tau^\mathrm{res}$ is the fraction of time the harmonic is resonant. Only harmonics with 
$\tau^\mathrm{res} > 1\%$ are shown.}
\footnotetext[3]{5th and 95th percentiles of the time distribution of $\halfwidth$ as subscripts and superscripts, respectively.}
\end{table}

\section{Quasi-integrals of motion}
\label{sec:quasi_integrals}
The emergence of a chaotic behavior of the planetary orbits can be explained in terms of the pendulum-like dynamics 
generated by each Fourier harmonic that composes the Hamiltonian in Eq.~\eqref{eq:hamiltonian} \citep[e.g.][]{Chirikov1979}. 
One can write $\Huv_{2n}(\vecI,\vectheta,t) = \fourcoeff{2n}{0}{0}(\vecI) + \sum_{i=1}^{\mathcal{M}_{2n}} \mathcal{F}_i(\vecI,\vectheta,t)$, 
with 
\begin{equation}
\label{eq:fourier_harmonic}
\F_i(\vecI,\vectheta,t) = 
\widetilde{\Huv}_{2n}^{\vec{k}^i\!,\vec{\ell}^i}\!(\vecI) \,
\E^{j \left( \vec{k}^i \cdot \vec{\theta} + \vec{\ell}^i \cdot \vec{\phi}(t) \right)}
+ \mathrm{c.c.} ,
\end{equation}
where $(\vec{k}^i,\vec{\ell}^i) \neq (\vec{0},\vec{0})$, $\mathcal{M}_{2n}$ is the number of harmonics in $\Huv_{2n}$ with a non-null 
wave vector, and $\mathrm{c.c.}$ stands for complex conjugate. 
Chaos arises from the interaction of resonant harmonics, that is, those harmonics $\mathcal{F}_i$ whose frequency 
combination $\vec{k}^i \cdot \dot{\vec{\theta}} + \vec{\ell}^i \cdot \dot{\vec{\phi}}(t)$ vanishes at some point along the motion. 
Using the computer algebra system \TRIP{}, the harmonics of $\Huv_{10}$ that enter into resonance along the 5-Gyr nominal solution of 
Gauss's dynamics have been systematically retrieved, together with the corresponding time statistics of the resonance 
half-widths $\halfwidth$ \citep{Mogavero2022}. 
The resonances have then been ordered by decreasing time median of their half-widths. The resulting ranking of resonances 
is denoted as $\R_1$ from now on. Table~\ref{tab:table1} recalls the 30 strongest resonances that are active for more than 
1\% of the 5-Gyr time span of the orbital solution. 
The wave vector of each harmonic is identified by the corresponding combination of frequency labels $(g_i, s_i)_{i=1}^8$, 
that is, $\veck \cdot \vecomegainn + \vecl \cdot \vecomegaout$, with $\vecomegainn = (g_1,g_2,g_3,g_4,s_1,s_2,s_3,s_4)$. 
Table~\ref{tab:table1} also shows the order of each harmonic, defined as the even integer $\O = \| (\veck, \vecl) \|_1$. 
The support of the asymptotic ensemble distribution of the FT-MLE shown in Fig.~\ref{fig:LCE4_1} overlaps in a robust way with that of 
the time distribution of the half-width of the strongest resonances. In other words, 
\begin{equation}
\label{eq:MLE_resonances}
2\pi \lambda_1 \approx \halfwidth^{\R_1} ,
\end{equation}
where $\halfwidth^{\R_1}$ stands for the half-width of the uppermost resonances of ranking $\R_1$. 
Equation~\eqref{eq:MLE_resonances} shows the dynamical sources of chaos in the ISS by connecting the top of the Lyapunov spectrum 
with the head of the resonance spectrum. Computer algebra allows us to establish such a connection in an unbiased way despite 
the multidimensional nature of the dynamics. 
We stress that such analysis is built on the idea that the time statistics of the resonant harmonics along 
a 5-Gyr ordinary orbital solution should be representative of their ensemble statistics 
(defined by a set of stable solutions with very close initial conditions) at some large time of the order 
of billions of years. This assumption was inspired by the good level of stationarity that characterizes the ensemble distribution 
of the MLE beyond 1 Gyr \cite{Mogavero2021,Mogavero2022}, and that extends to the entire spectrum in Fig.~\ref{fig:LCE4_1}. 

We remark that, strictly speaking, ranking $\R_1$ is established on the Fourier harmonics of the Lie-transformed Hamiltonian 
$\Huv_{2n}^\prime$ \citep[][Appendix~G]{Mogavero2022}. New canonical variables are indeed defined to transform $\Huv_{2n}$ in 
a Birkhoff normal form to degree 4. The goal is to let the interactions of the terms of degree 4 in $\Huv_{2n}$ appear 
more explicitly in the amplitudes of the harmonics of higher degrees in $\Huv_{2n}^\prime$, the physical motivation being that the 
non-linear interaction of the harmonics at degree 4 constitutes the primary source of chaos \citep{Mogavero2022}. Keeping in 
mind the quasi-identity nature of the Lie transform, here we drop for simplicity the difference between the two Hamiltonians. 
Moreover, all the new analyses of this work involve the original variables of Eq.~\eqref{eq:AAvars}. 

\subsection{Quasi-symmetries of the resonant harmonics}
\label{sec:quasi_symmetries}
In addition to the dynamical interactions responsible for the chaotic behavior of the orbits, Table~\ref{tab:table1} provides 
information on the geometry of the dynamics in the action variable space. Ranking the Fourier harmonics allows us to consider 
partial Hamiltonians constructed from a limited number $m$ of leading terms \citep{Mogavero2022,Hoang2022}, that is, 
\begin{equation}
\label{eq:partial_ham}
\Huv_{2n,m} = 
\fourcoeff{2n}{0}{0}
+ \textstyle{\sum_{i=1}^m} \F_i .
\end{equation}
The dynamics of a Hamiltonian reduced to a small set of harmonics is generally characterized by several symmetries and corresponding 
integrals of motion. We are interested in how these symmetries are progressively destroyed when one increases the number of terms 
taken into account in Eq.~\eqref{eq:partial_ham}. 

Consider a set of $m$ harmonics of $\Huv_{2n}$ and a dynamical quantity that is a linear combination of the action variables, that is, 
\begin{equation}
\label{eq:QI_def}
C_{\vecgamma} = \vecgamma \cdot \vecI , 
\end{equation}
$\vecgamma \in \mathbb{R}^8$ being a parameter vector. From Eq.~\eqref{eq:fourier_harmonic}, the partial contribution of the $m$ harmonics 
to the time derivative of $C_{\vecgamma}$ along the flow of $\Huv_{2n}$ is 
\begin{equation}
\label{eq:Q_deriv}
\dot{C}_{\vecgamma,m} = 2 \sum_{i=1}^m \vecgamma \cdot \veck^i 
\operatorname{Im} \{
\widetilde{\Huv}_{2n}^{\vec{k}^i\!,\vec{\ell}^i}\!(\vecI) \,
\E^{j \left( \vec{k}^i \cdot \vec{\theta} + \vec{\ell}^i \cdot \vec{\phi}(t) \right)}
\} ,
\end{equation}
and $\dot{C}_{\vecgamma} = \dot{C}_{\vecgamma,\mathcal{M}_{2n}}$, where $\mathcal{M}_{2n}$ is the total number 
of harmonics with a non-null wave vector that appear in $\Huv_{2n}$. 
Any quantity $C_{\vecgamma}$ with $\vecgamma \cdot \veck^i = 0$ is conserved by the one-d.o.f. dynamics generated by 
the single harmonic $\F_i$. In other words, such a quantity would be an integral of motion if $\F_i$ were the only harmonic to appear in the 
Hamiltonian. Considering now $m$ different harmonics, these do not contribute to the change of the quantity $C_{\vecgamma}$ 
if $\vecgamma \perp \Span(\veck^1, \veck^2, \dots, \veck^m)$, that is, if the vector $\vecgamma$ belongs to the orthogonal complement 
of the linear subspace of $\mathbb{R}^8$ spanned by the wave vectors $(\veck^i)_{i=1}^m$. 
We also consider the quantity 
\begin{equation}
\label{eq:QI_def2}
C_{\vecgamma}^\prime = \Huv_{2n} + \vecgamma \cdot \vecI . 
\end{equation}
Because of the explicit time dependence in the Hamiltonian, the partial contribution of a set of $m$ harmonics 
to the time derivative of $C_{\vecgamma}^\prime$ along the flow of $\Huv_{2n}$ is 
\begin{equation}
\label{eq:Qprime_deriv}
\dot{C}^\prime_{\vecgamma,m} = 2 \sum_{i=1}^m (\vecgamma \cdot \veck^i + \vecl^i \cdot \vecomegaout) 
\operatorname{Im} \{
\widetilde{\Huv}_{2n}^{\vec{k}^i\!,\vec{\ell}^i}\!(\vecI) \,
\E^{j \left( \vec{k}^i \cdot \vec{\theta} + \vec{\ell}^i \cdot \vec{\phi}(t) \right)}
\} ,
\end{equation}
and one has $\dot{C}^\prime_{\vecgamma} = \dot{C}^\prime_{\vecgamma,\mathcal{M}_{2n}}$. Quantity $C_{\vecgamma}^\prime$ is unchanged by 
the $m$ harmonics if $\vecgamma \cdot \veck^i + \vecl^i \cdot \vecomegaout = 0$ for $i \in \{1,2,\dots,m\}$. 
Dynamical quantities $C_{\vecgamma}$ or $C^\prime_{\vecgamma}$ that are unaffected by a given set of leading harmonics, that is, 
with null partial contribution in Eq.~\eqref{eq:Q_deriv} or \eqref{eq:Qprime_deriv}, are denoted as quasi-integrals of motion 
from now on. More specifically, we build our analysis on ranking $\R_1$, since the resonant harmonics are those responsible 
for changes that accumulate stochastically over long timescales, driving chaotic diffusion. 

In the framework of the aforementioned considerations, the resonances listed in Table~\ref{tab:table1} possess three different 
symmetries.
\paragraph{First symmetry.} 
The rotational invariance of the entire Solar System implies the d'Alembert rule 
$\sum_{i=1}^8 k_{i} + \sum_{i=1}^7 \ell_{i} = 0$, where $\veck = (k_1,k_2,\dots,k_8)$ and $\vecl = (\ell_1,\ell_2,\dots,\ell_7)$ 
\citep{Laskar1990c,Laskar1995,Morbidelli2002,Boue2009}. Moreover, the Jupiter-dominated eccentricity mode $g_5$ is the only fundamental 
Fourier mode of the outer planet forcing to appear in Table~\ref{tab:table1}. The quantity 
\begin{equation}
\label{eq:energy}
\mathcal{E}_{2n} \vcentcolon = C_{g_5 \vec{1}_8}^\prime = \Huv_{2n} + g_5 \textstyle{\sum_{i=1}^4} (\Chi_i + \Psi_i),
\end{equation}
with $\vec{1}_8 = (1,1,\dots,1) \in \mathbb{R}^8$, is therefore unaffected by the resonances listed in Table~\ref{tab:table1}. 
In an equivalent way, the time-dependent canonical transformation 
$\vectheta \rightarrow \vectheta + g_5 t \, \vec{1}_8$, with unchanged action variables, allows us to 
remove the explicit time dependence in these harmonics. Quantity $\mathcal{E}_{2n}$ coincides with the 
transformed Hamiltonian and the harmonics in Table~\ref{tab:table1} do not contribute to its time derivative. 
\paragraph{Second symmetry.} 
We write the eccentricity and inclination parts of the harmonic wave vectors explicitly, that is, $\veck = (\vecke,\vecki)$ 
with $\vecke,\vecki \in \mathbb{R}^4$. 
One can visually check that the harmonics in Table~\ref{tab:table1} verify the relation $\sum_{i=1}^4 \superscript{k}{inc}_i = 0$, where 
$\vecki~=~(\superscript{k}{inc}_1,\dots,\superscript{k}{inc}_4)$. Therefore, denoting $\vec{\gamma}_1 = (\vec{0}_4, \vec{1}_4)$, the quantity 
\begin{equation}
\label{eq:Cinc}
\Cinc \vcentcolon = C_{\vec{\gamma}_1} = \Psi_1 + \Psi_2 + \Psi_3 + \Psi_4
\end{equation}
is conserved by these resonances. $\Cinc$ is the angular momentum deficit (AMD) \citep{Laskar1997} contained 
in the inclination d.o.f. This symmetry can then be interpreted as a remnant of the conservation of the AMD of the 
entire (secular) Solar System. We remark that the AMD contained in the eccentricity d.o.f., $\Cecc = \sum_{i=1}^4 \Chi_i$, 
is not invariant under the leading resonances because of the eccentricity forcing mainly exerted by Jupiter through 
the mode $g_5$. 
The conservation of $\Cinc$ depends on two facts: the inclination modes $s_6,s_7,s_8$ of the external forcing do not 
appear in Table~\ref{tab:table1}; low-order harmonics like $2g_1-s_1-s_2$, $2g_1-2s_1$, and $2g_1-2s_2$ are never 
resonant (even if they can raise large quasi-periodic contributions), so that two AMD reservoirs $\Cecc$ and $\Cinc$ 
are decoupled in Table~\ref{tab:table1}. We recall that the absence of an inclination mode $s_5$ in the external 
forcing relates to the fixed direction of the angular momentum of the entire Solar System \citep{Laskar1990,Morbidelli2002,Boue2009}. 

\paragraph{Third symmetry.} 
The first two symmetries could be expected to some extent on the basis of physical intuition of the interaction between outer and 
inner planets. 
However, it is not easy to even visually guess the third one from Table~\ref{tab:table1}. Consider the $30\times8$ matrix 
$\mathbf{K}_{30}$ whose rows are the wave vectors $(\veck^i)_{i=1}^{30}$ of the listed resonances. A singular value decomposition 
shows that the rank of $\mathbf{K}_{30}$ is equal to 6. Therefore, the linear subspace 
$V_{30} = \Span(\veck^1, \veck^2, \dots, \veck^{30})$ spanned by the wave vectors has dimension 6. 
A Gram–Schmidt orthogonalization allows us to determine two linearly independent vectors that span its 
orthogonal complement $V_{30}^\perp$. One choice consists in $V_{30}^\perp = \Span(\vecgamma_2,\vecgamma_2^\perp)$, with 
\begin{equation}
\label{eq:C4_C5}
\begin{aligned}
\C_2 \vcentcolon= \C_{\vecgamma_2} &= -\Chi_3 - \Chi_4 + \Psi_1 + \Psi_2 + 2\Psi_3 + 2\Psi_4 , \\
\C_2^\perp \vcentcolon= \C_{\vecgamma_2^\perp} &= \phantom{-}\Chi_3 + \Chi_4 + \Psi_1 + \Psi_2 .
\end{aligned}
\end{equation}
Since the second symmetry clearly requires that $\vec{\gamma}_1 \in V_{30}^\perp$, the three quantities 
$\Cinc, \C_2, \C_2^\perp$ are not independent and one has indeed $\Cinc = (\C_2 + \C_2^\perp)/2$. We remark that 
$(\C_2 - \C_2^\perp)/2 = -\Chi_3 - \Chi_4 + \Psi_3 + \Psi_4$. The additional symmetry can thus be interpreted in terms 
of a certain decoupling between the d.o.f. 3,\,4 and 1,\,2, representing in the proper modes the Earth-Mars and Mercury-Venus 
subsystems, respectively. 

The aforementioned symmetries, that exactly characterize the resonances listed in Table~\ref{tab:table1}, 
naturally represent quasi-symmetries when considering the entire spectrum of resonances $\R_1$. They are indeed broken 
at some point by weak resonances (see Sect.~\ref{sec:secondary_resonances}). Quantities $\mathcal{E}_{2n}$, $\Cinc$, and $\C_2$ 
are the corresponding QIs of motion. 
The persistence of the three symmetries under the 30 leading resonances is somewhat surprising. 
Concerning $\Cinc$ and $\C_2$, for example, one might reasonably expect that, since the ISS has 8 d.o.f., 
the subspace spanned by the wave vectors of just a dozen of harmonics should already have maximal dimension, 
destroying all possible symmetries. 

We remark that, differently from $\Cinc$ and $\C_2$, the quantity $\mathcal{E}_{2n}$ is a non-linear function of 
the action-angle variables. However, as far as stable orbital evolutions are concerned, the convergence 
of the series expansion of the Hamiltonian is sufficiently fast that the linear LL approximation 
$\mathcal{E}_2 = \Huv_2 + g_5 \vec{1}_8 \cdot \vecI = C_{\vec{\gamma}_3}$, 
with $\vec{\gamma}_3 = -\LL{\vec{\omega}} + g_5 \vec{1}_8$, reproduces reasonably well $\mathcal{E}_{2n}$ 
along the flow of $\Huv_{2n}$ for $n > 1$. The vector $\vec{\gamma}_3$ is used in Sect.~\ref{sec:PCA}, 
together with $\vec{\gamma}_1$ and $\vec{\gamma}_2$, to deal with the geometry of the linear action subspace spanned 
by the QIs. The explicit expressions of these vectors are given in Appendix~\ref{sec:gamma_vectors}. We mention that, 
differently from $\vec{\gamma}_1$ and $\vec{\gamma}_2$, the components of $\vec{\gamma}_3$ are not integer and 
they have the dimension of a frequency. 

\begin{figure*}
\includegraphics[width=\textwidth]{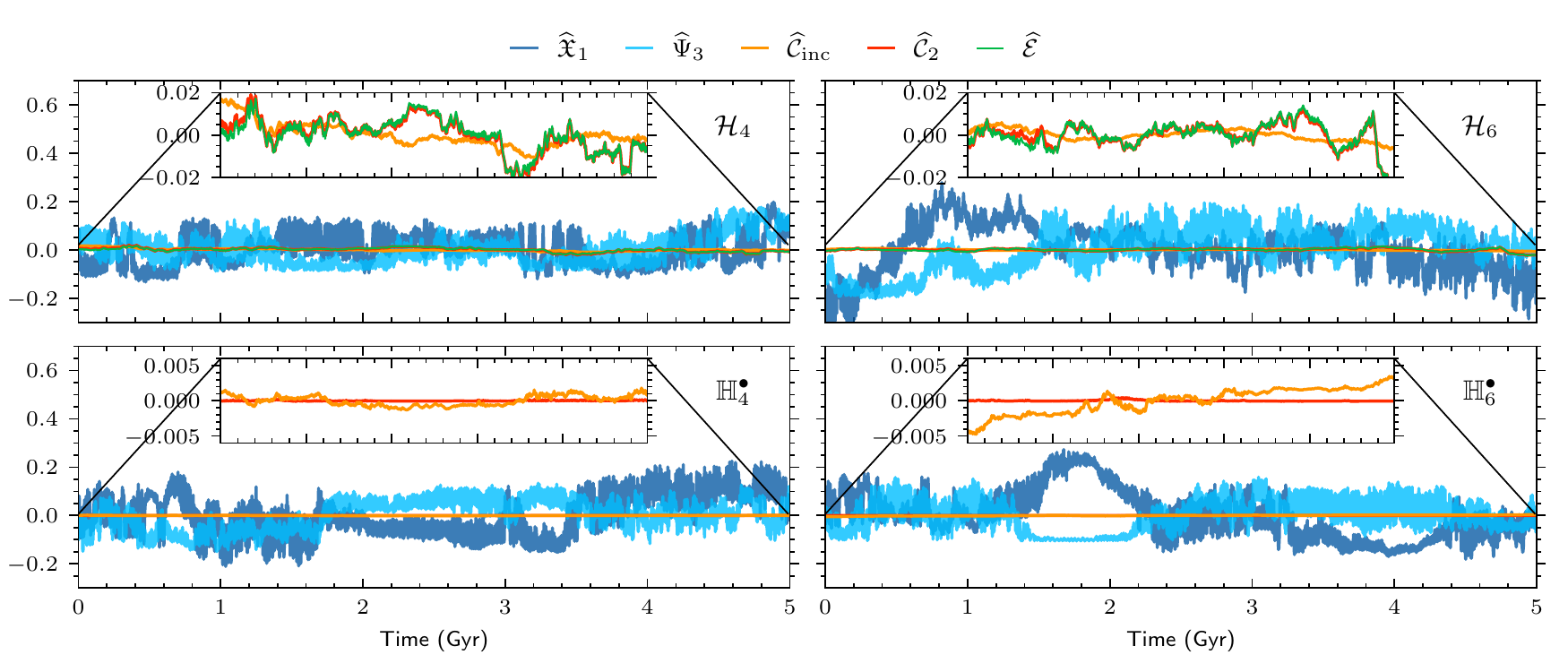}
\caption{\label{fig:slowfast}Time evolution over 5 Gyr of the dimensionless QIs 
($\Cinchat, \widehat{\mathcal{C}}_2, \widehat{\mathcal{E}}$) and of two 
representatives of the dimensionless action variables ($\widehat{\Chi}_1, \widehat{\Psi}_3$) along the nominal orbital solutions 
of different models. \textit{Top row}: $\Hiss_4$ and $\Hiss_6$ ($\widehat{\mathcal{E}}$ stands for 
$\widehat{\mathcal{E}}_4$ and $\widehat{\mathcal{E}}_6$, respectively). \textit{Bottom row}: $\simpleH{4}$ and $\simpleH{6}$ 
from Eq.~\eqref{eq:simpleH} ($\widehat{\mathcal{E}}_{2n}^\bullet$ is exactly conserved and not shown). 
The time series are low-pass filtered with a cutoff frequency of 1 Myr$^{-1}$ and the mean over 5 Gyr is subtracted. 
The variations of the QIs are enlarged in the insets.
The $\simpleH{2n}$ models are introduced and discussed in Sec.~\ref{sec:new_truncation}.}
\end{figure*}

\subsection{Slow variables}
\label{sec:numerical_checks}
The QIs of motion $\mathcal{E}_{2n},\Cinc,\C_2$ are clearly strong candidates for slow variables once evaluated 
along the orbital solutions. In what follows, to assess the slowness of a dynamical quantity when compared to the 
typical variations of the action variables, we consider the variance of its time series along a numerical solution. 

We define the dimensionless QIs 
\begin{equation}
\label{eq:dimensionless_QIs}
\Cinchat = \frac{\Cinc}{\Vert \vec{\gamma}_1 \Vert \textrm{C}_0}, \quad
\widehat{\C}_2 = \frac{\C_2}{\Vert \vec{\gamma}_2 \Vert \textrm{C}_0}, \quad
\widehat{\mathcal{E}}_{2n} = \frac{\mathcal{E}_{2n}}{\Vert \vec{\gamma}_3 \Vert \textrm{C}_0}, 
\end{equation} 
where $\textrm{C}_0$ stands for the current total AMD of the inner planets, that is, the value of $\Cecc + \Cinc$ at time zero. 
We stress that, by introducing the unit vectors $\widehat{\vec{\gamma}}_i = \vec{\gamma}_i / \Vert \vec{\gamma}_i \Vert$ 
for $i \in \{1,2,3\}$, one has $\Cinchat = C_{\widehat{\vec{\gamma}}_1} / \textrm{C}_0$ and 
$\widehat{\C}_2 = C_{\widehat{\vec{\gamma}}_2} / \textrm{C}_0$. At degree 2, one also has 
$\widehat{\mathcal{E}}_2 = C_{\widehat{\vec{\gamma}}_3}/\textrm{C}_0$. 
We then consider the ensembles of numerical integrations of $\Hiss_4$ and $\Hiss_6$, with very close initial conditions 
and spanning 100 Gyr in the future, that have been presented in Ref.~\citep{Hoang2022}. 
The top row of Fig.~\ref{fig:slowfast} shows the time evolution over 5 Gyr of the dimensionless QIs and of two components 
of the dimensionless action vector $\widehat{\vecI} = \vecI / \textrm{C}_0$ along the nominal orbital solutions 
of the two ensembles. We subtract from each time series its mean over the plotted time span. 
The time series are low-pass filtered by employing the Kolmogorov-Zurbenko (KZ) filter 
with three iterations of the moving average \citep{Zurbenko2017,Mogavero2021}. A cutoff frequency of 1 Myr$^{-1}$ is chosen 
to highlight the long-term diffusion that can be hidden by short-time quasi-periodic oscillations. This is in line with our 
definition of quasi-integrals based on contribution from resonant harmonics only. Figure~\ref{fig:slowfast} clearly shows 
that the QIs are slowly-diffusing variables when compared to an arbitrary function of the action variables. 
The behavior of the QIs along the nominal orbital solutions of Fig.~\ref{fig:slowfast} is confirmed by a statistical analysis 
in Appendix~\ref{sec:ensemble_stats_QIs}. Figure~\ref{fig:PDF_QIs} shows the time evolution of the distributions of 
the same quantities as Fig.~\ref{fig:slowfast} over the stable orbital solutions of the entire ensembles of 1080 numerical integrations 
of Ref.~\citep{Hoang2022}. Figure~\ref{fig:IQR_QIs} details the growth of the QI dispersion over time. 

We remark that $\C_2$ and $\mathcal{E}_{2n}$ show very similar time evolutions along stable orbital solutions, 
as can be seen in the top row Fig.~\ref{fig:slowfast}. This is explained by the interesting observation that the components of 
the unit vectors $\widehat{\vec{\gamma}}_2$ and $\widehat{\vec{\gamma}}_3$ differ from each other by only a few percent, as 
shown in Appendix~\ref{sec:gamma_vectors}. However, we stress that the two vectors are in fact linearly independent: $\C_2$ 
does not depend on the actions $\Chi_1$ and $\Chi_2$, while $\mathcal{E}_{2n}$ does. The two QIs move away from 
each other when high eccentricities of Mercury are reached, that is, for large excursions of the Mercury-dominated action $\Chi_1$. 

\subsection{Weak resonances and Lyapunov spectrum}
\label{sec:secondary_resonances}
A fundamental result from Table~\ref{tab:table1} is that the symmetries introduced in Sect.~\ref{sec:quasi_symmetries} are still 
preserved by resonances that have half-widths an order of magnitude smaller than those of the strongest terms. It is 
natural to extract from ranking $\R_1$ the weak resonances that break the three symmetries. A new ranking of resonances $\R_2$ 
is defined in this way. Table~\ref{tab:table2} reports the 10 strongest symmetry-breaking resonances that change 
$\mathcal{E}_{2n},\Cinc,\C_2$, respectively. As in Table~\ref{tab:table1}, only harmonics that are resonant for more than 
1\% of the 5-Gyr time span of the nominal solution of Gauss's dynamics are shown. 
The leading symmetry-breaking resonances have half-widths of about 0.01\arcsecyr{}. For each QI, the dominant contribution comes 
from harmonics involving Fourier modes of the outer planet forcing other than $g_5$: the Saturn-dominated modes $g_6,s_6$ 
and the modes $g_7,s_7$ mainly associated to Uranus. In the case of $\Cinc$, there is also a contribution that starts 
at about 0.006\arcsecyr{} with $\F_8 = 4g_1 - g_2 - g_3 - s_1 - 2s_2 + s_4$ and comes from high-order internal resonances, 
that is, resonances that involve only the d.o.f. of the inner planets. 
We remark that the decrease of the resonance half-width with the index of the harmonic in Table~\ref{tab:table2} is steeper 
for $\Cinc$ than for $\mathcal{E}_{2n},\C_2$, and is accompanied by a greater presence of high-order resonances. This may 
notably explain why the secular variations of $\Cinc$ are somewhat smaller in the top row of Fig.~\ref{fig:slowfast}. 
We finally point out the important symmetry-breaking role of the modes $g_7,s_7$, representing the forcing mainly exerted by Uranus. 
Differently from what one might suppose, these modes cannot be completely neglected when addressing the long-term diffusion of ISS. 
This recalls the role of the modes $s_7$ and $s_8$ in the spin dynamics of Venus \citep{Correia2003}, and is basically a 
manifestation of the long-range nature of the gravitational interaction. 

\begin{table}
\caption{\label{tab:table2}
Top of ranking $\R_2$. 
First 10 symmetry-breaking resonances of $\Huv_{10}$ along the 5-Gyr nominal solution of Gauss's dynamics, 
that change $\mathcal{E}_{2n}$, $\Cinc$, and $\C_2$, respectively (see Table~\ref{tab:table1} for details).} 
\begin{ruledtabular}
\begin{tabular}{rrrrc}
$i$ & Fourier harmonic $\F_i$ & $\O_i$ & $\tau_i^\mathrm{res}$ & $\halfwidth_i$ \\
\colrule
\multicolumn{5}{c}{$\mathcal{E}_{2n}$} \\
\colrule
\rule{0pt}{2.3ex}
\input{tables/resonances_energy_all_modes.out}
\colrule
\multicolumn{5}{c}{$\Cinc$} \\
\colrule
\rule{0pt}{2.3ex}
\input{tables/resonances_Cinc_all_modes.out}
\colrule
\multicolumn{5}{c}{$\C_2$} \\
\colrule
\rule{0pt}{2.3ex}
\input{tables/resonances_C4_all_modes.out}
\end{tabular}
\end{ruledtabular}
\end{table}

\begin{table}
\caption{\label{tab:table3}
Top of ranking $\R_3$. 
First 10 symmetry-breaking resonances of $\Huv_{10}$ along the 5-Gyr nominal solution of Gauss's dynamics, 
that only involve $g_5$ among the external modes and change $\mathcal{E}_{2n}$, $\Cinc$, and $\C_2$, respectively.} 
\begin{ruledtabular}
\begin{tabular}{rrrrc}
$i$ & Fourier harmonic $\F_i$ & $\O_i$ & $\tau_i^\mathrm{res}$ & $\halfwidth_i$ \\
\colrule
\multicolumn{5}{c}{$\mathcal{E}_{2n}$} \\
\colrule
\\
\multicolumn{5}{c}{No resonances} \\
\\
\colrule
\multicolumn{5}{c}{$\Cinc$\footnotemark[1]} \\
\colrule
\rule{0pt}{2.3ex}
\input{tables/resonances_Cinc_g5.out}
\colrule
\multicolumn{5}{c}{$\C_2$} \\
\colrule
\rule{0pt}{2.3ex}
\input{tables/resonances_C4_g5.out}
\end{tabular}
\end{ruledtabular}
\footnotetext[1]{Only harmonics that are resonant for more than one percent of time are shown, i.e., $\tau^\mathrm{res} > 1\%$.}
\end{table}

As we state in Sect.~\ref{sec:spectrum}, a pair of Lyapunov exponents would vanish if there were an exact integral of motion. 
In the presence of a weakly broken symmetry, one may expect a small positive Lyapunov exponent whose value relates to the 
half-width of the strongest resonances driving the time variation of the corresponding QI. Such an argument is a natural extension 
of the correspondence between the FT-MLE and the top of the resonance spectrum given in Eq.~\eqref{eq:MLE_resonances}. 
Comparison of Table~\ref{tab:table2} with the Lyapunov spectrum in Fig.~\ref{fig:LCE4_1} shows that the time statistics 
of the half-widths of the symmetry-breaking resonances of ranking $\R_2$ overlaps with the ensemble distribution of the 
three smallest FT-LCEs, that is, $\lambda_6, \lambda_7, \lambda_8$. One can indeed write 
\begin{equation}
\label{eq:FLEs_resonances}
2\pi \lambda_6 \approx \halfwidth^{\R_2} ,
\end{equation}
where $\halfwidth^{\R_2}$ stands for the half-width of the uppermost resonances of ranking $\R_2$. Table~\ref{tab:table2} 
and Fig.~\ref{fig:LCE4_1} suggest a relation between the QIs and the smallest Lyapunov exponents: 
\begin{equation}
\label{eq:FLEs_QIs}
\lambda_6, \lambda_7, \lambda_8 \longleftrightarrow \mathcal{E}_{2n},\Cinc,\C_2 .
\end{equation}
Equation~\eqref{eq:FLEs_QIs} is not a one-to-one correspondence, nor should it be understood as an exact relation since, 
for example, $\lambda_6$ is not well separated from the larger exponents. Its physical meaning is that the QIs are among 
the slowest d.o.f. of the ISS dynamics. Such a claim is one of the core points of this work. 
In Sect.~\ref{sec:PCA}, we show its statistical validity in the geometric framework established by a principal component analysis of the 
orbital solutions. Moreover, Sect.~\ref{sec:new_truncation} shows that Eq.~\eqref{eq:FLEs_QIs} can be stated more precisely 
in the case of a simplified dynamics that underlies $\Huv_{2n}$.
We remark that $\mathcal{E}_{2n},\Cinc,\C_2$ constitute a set of three QIs that are independent and \textit{nearly in involution}, 
and it is thus meaningful to associate three different Lyapunov exponents with them. On the one hand, the independence is easily checked at 
degree 2 as the vectors $\vec{\gamma}_1, \vec{\gamma}_2, \vec{\gamma}_3$ are linearly independent. On the other hand, one has 
the Poisson bracket $\{\Cinc,\C_2\} = 0$, since the two quantities are functions of the action variables only. One also has 
$\{\mathcal{E}_{2n},\Cinc\} = \{\Huv_{2n},\Cinc\} = \subscript{\dot{\mathcal{C}}}{inc}$ and $\{\mathcal{E}_{2n},\C_2\} = \dot{\C}_2$. 
Only weak resonances contribute to these Poisson brackets and the three QIs are therefore nearly in involution. 

\subsection{New truncation of the Hamiltonian}
\label{sec:new_truncation}
The fundamental role of the external modes $g_6,g_7,s_6,s_7$ in Table~\ref{tab:table2} raises the question of which symmetry-breaking 
resonances persist if one excludes all the Fourier harmonics that involve external modes other than $g_5$. Therefore, we define a new 
ranking $\R_3$ by extracting such resonances from ranking $\R_2$. Table~\ref{tab:table3} reports the 10 strongest resonances per each 
broken symmetry. The difference with respect to Table~\ref{tab:table2} is manifest. As $g_5$ is the only external mode remaining, 
there are no resonances left that can contribute to the time evolution of $\mathcal{E}_{2n}$. For the remaining two QIs, 
the only harmonics that appear in Table~\ref{tab:table3} are of order 8 or higher, and this is accompanied by a significant drop in 
the half-width of the leading resonances. 
In the case of $\Cinc$, the half-width of the uppermost resonances is now around 0.005\arcsecyr{}. One can appreciate that 
the activation times $\tau^\mathrm{res}$ of the resonances do not exceed a few percent, differently from Table~\ref{tab:table2}. 
The most impressive change is, however, related to $\C_2$: only harmonics of order 10 appear in Table~\ref{tab:table3} and the 
half-width of the uppermost resonances drops by two orders of magnitude. We stress that such harmonics are resonant for 
very short periods of time along the 5 Gyr spanned by the nominal solution of Gauss's dynamics. To retrieve the time statistics 
of the resonances affecting $\C_2$, we indeed choose to repeat the computations of Ref.~\citep{Mogavero2022} by increasing 
the cutoff frequency of the low-pass filter applied to time series of the action-angle variables from (5 Myr)$^{-1}$ to 1 Myr$^{-1}$ 
\citep[][Appendices~F.2 and G.5]{Mogavero2022}. The filtered time series have then been resampled with a timestep of 50 kyr. 
Many harmonics we show in Table~\ref{tab:table3} and related to $\C_2$ are resonant for a few timesteps and their time statistics is 
very tentative. More precise estimations of the half-widths should be obtained over an ensemble of different orbital solutions, 
possibly spanning more than 5 Gyr. In any case, the fundamental point here is the drastic reduction in the size of the uppermost resonances 
with respect to Table~\ref{tab:table2}, and this is a robust result. We remark that resonances of order 12 and 
higher may also carry an important contribution at these scales, but they are excluded by the truncation at degree 10 adopted 
in Ref.~\citep{Mogavero2022} to establish the resonant harmonics, so they do not appear in the tables of this work. 

\paragraph*{Hamiltonian $\simpleH{2n}$.} 
The implications of Table~\ref{tab:table3} suggest to introduce an additional truncation in the Hamiltonian $\Huv_{2n}$. 
This consists in dropping the harmonics of Eq.~\eqref{eq:hamiltonian} that involve external modes other than $g_5$: 
\begin{equation}
\label{eq:simpleH}
\simpleH{2n}(\vecI,\vectheta,t) = 
\sum_{\substack{\veck,\ell_1}}
\widetilde{\Huv}_{2n}^{\veck,\vecl^\bullet} \! (\vecI)
\E^{j \left( \vec{k} \cdot \vec{\theta} + \ell_1 \phi_1(t) \right)} ,
\end{equation}
where $\phi_1(t) = -g_5 t$ and $\vecl^\bullet = (\ell_1,0,\dots,0)$, with $\ell_1 \in \mathbb{Z}$. 
Consistently with the absence of symmetry-breaking resonances related to $\mathcal{E}_{2n}$ in Table~\ref{tab:table3}, the 
corresponding dynamics admits the exact integral of motion 
\begin{equation}
\label{eq:energy_bullet}
\mathcal{E}_{2n}^\bullet = \simpleH{2n} + g_5 \textstyle{\sum_{i=1}^4} (\Chi_i + \Psi_i) ,
\end{equation} 
which represents the transformed 
Hamiltonian under the canonical change of variables that eliminates the explicit time dependence in Eq.~\eqref{eq:simpleH}. 
We point out that, as the additional truncation is applied to the action-angle formulation of Eq.~\eqref{eq:hamiltonian}, the 
external modes other than $g_5$ still enter the definition of the proper modes of the forced Laplace-Lagrange dynamics 
\citep{Mogavero2021}. 
The orbital solution arising from $\simpleH{2n}$ is initially very close to that of $\Huv_{2n}$. A frequency analysis 
over the first 20 Myr shows that the differences in the fundamental frequencies of the motion between $\simpleH{2n}$ and 
$\Huv_{2n}$ are of the order of $10^{-3}$ \arcsecyrtext{}, an order of magnitude smaller than the typical frequency differences between $\Hiss_4$ 
and $\Hiss_6$ \citep[][Table~3]{Mogavero2021}. Therefore, even though $\simpleH{2n}$ constitutes a simplification of $\Huv_{2n}$, 
it should not be regarded as a toy model. Its dynamics, in particular, still possesses 8 d.o.f. 

We compute the Lyapunov spectrum of the Hamiltonian $\simpleH{4}$ in the same way as described in Sect.~\ref{sec:spectrum} 
in the case of $\Hiss_{2n}$. Since its dynamics turns out to be much more stable than that of $\Hiss_4$ (see 
Sect.~\ref{sec:implications}, Fig.~\ref{fig:CDFs}), we extend the computation to a time span of 100 Gyr. The marginal 
ensemble PDFs of the positive FT-LCEs are shown in Fig~\ref{fig:LCE4_2}. Comparing to the Lyapunov spectrum of $\Hiss_4$, 
one notices that the distributions of the leading exponents turn out to be quite similar, apart from being more spaced and 
except for a slight decrease in their median values. However, such a decrease is more pronounced for smaller exponents, and 
the drop in the smallest exponents is drastic. The smallest one, $\lambda_8$, decreases monotonically, consistently with the fact 
that $\mathcal{E}^\bullet_4$ from Eq.~\eqref{eq:energy_bullet} is an exact integral of motion. The exponent $\lambda_7$ drops 
by more than an order of magnitude, and apparently begins to stabilize around a few $10^{-4}$ \arcsecyrtext{}, while $\lambda_6$ 
also reduces significantly, by a factor of three, to about 0.005\arcsecyr{}. 
The drop in the smallest exponents agrees remarkably well with that of the half-width of the leading symmetry-breaking resonances 
when switching from Table~\ref{tab:table2} to Table~\ref{tab:table3}. One can indeed write 
\begin{equation}
\label{eq:FLEs_resonances_simpleH}
\begin{aligned}
2\pi \lambda_6 &\approx \halfwidth^{\R_3,\Cinc}, \\
2\pi \lambda_7 &\approx \halfwidth^{\R_3,\C_2}, \\
\lambda_8 &= 0 , 
\end{aligned}
\end{equation}
where $\halfwidth^{\R_3,Q}$ stands for the half-width of the uppermost resonances of ranking $\R_3$ related to the quasi-integral $Q$. 
The hierarchy of the three smallest exponents in the spectrum of Fig.~\ref{fig:LCE4_2} consistently follows that of the QIs suggested 
in Table~\ref{tab:table3} by the very different sizes of the leading resonances. In other words, one can state: 
\begin{equation}
\label{eq:hierarchy_simpleH}
\begin{aligned}
\lambda_6 &\longleftrightarrow \Cinc, \\
\lambda_7 &\longleftrightarrow \C_2, \\
\lambda_8 &\longleftrightarrow \mathcal{E}_{2n}^\bullet . 
\end{aligned}
\end{equation}
These one-to-one correspondences are a particular case of Eq.~\eqref{eq:FLEs_QIs} and support the physical intuition behind it. 
In Sect.~\ref{sec:PCA}, we prove the validity of Eq.~\eqref{eq:hierarchy_simpleH} in the geometric framework established 
by a principal component analysis of the orbital solutions of $\simpleH{2n}$. 

\paragraph*{Numerical integrations.} 
We compute ensembles of 1080 orbital solutions of the dynamical models $\simpleH{4}$ and $\simpleH{6}$, with initial conditions 
very close to the nominal ones of Gauss's dynamics and spanning 100 Gyr in the future. This closely follows what we did in 
Ref.~\citep{Hoang2022} in the case of the models $\Hiss_{2n}$. The bottom row of Fig.~\ref{fig:slowfast} shows the filtered 
dimensionless QIs along the nominal solutions of the two models over the first 5 Gyr. The hierarchy of the QIs stated in 
Eq.~\eqref{eq:hierarchy_simpleH} is manifest. The quantity $\C_2$ has secular variations 
much slower than $\Cinc$, while the latter is itself slower with respect to its counterpart in the orbital solutions of 
$\Hiss_{2n}$. We remark that, as $\mathcal{E}_{2n}^\bullet$ is an exact integral of motion for the model $\simpleH{2n}$, 
we do not plot it. From Fig.~\ref{fig:slowfast} it is also evident how difficult can be the retrieval of the short-lasting resonances 
affecting $\C_2$ from a solution of $\simpleH{2n}$ spanning only a few billion years. 

The hierarchy of the QIs is confirmed by a statistical analysis in Appendix~\ref{sec:ensemble_stats_QIs}. Figure~\ref{fig:PDF_QIs} shows 
the entire time evolution of the distributions of the filtered dimensionless QIs over the stable orbital solutions of the ensembles of 
1080 numerical integrations. Figure~\ref{fig:IQR_QIs} details the growth of the QI dispersion over time. As suggested by 
Table~\ref{tab:table3}, the drop in the diffusion rates of the QIs when switching from $\Hiss_{2n}$ to $\simpleH{2n}$ is manifest. 

\section{Statistical detection of \\slow variables}
\label{sec:PCA}
Section~\ref{sec:quasi_integrals} shows how the slow-fast nature of the ISS dynamics, indicated by the Lyapunov spectrum, emerges 
from the quasi-symmetries of the resonant harmonics of the Hamiltonian. QIs of motion can be introduced semi-analytically 
and they constitute slow quantities when evaluated along stable orbital solutions. 
In this section, we consider the slow variables that can be systematically retrieved from a numerically integrated orbital 
solution by means of a statistical technique, the principal component analysis. We show that, in the case of the 
forced secular ISS, the slowest variables are remarkably close to the QIs, and this can be established in a precise geometric 
framework. 

\subsection{Principal component analysis}
PCA is a widely used classical technique for multivariate analysis \cite{pearson1901,hotelling1933}.
For a given dataset, PCA aims to find an orthogonal linear transformation of the variables such that the new coordinates 
offer a more condensed and representative view of the data. The new variables are called principal components (PCs). They are uncorrelated 
and ordered according to decreasing variance: the first PC and last one have, respectively, the largest and the smallest variance 
of any linear combination of the original variables. While one is typically interested in the PCs of largest variance, in this work 
we employ the variance of the time series of a dynamical quantity to assess its slowness when compared to the typical variations 
of the action variables (see Sect.~\ref{sec:numerical_checks}). We thus perform a PCA of the action variables $\vecI$ and 
focus on the last PCs, as they give a pertinent statistical definition of slow variables. We stress that, when coupled to a 
low-pass filtering of the time series, the statistical variance provides a measure of chaotic diffusion. 

\paragraph*{Implementation.}
Our procedure for the PCA is described briefly as follows \citep[for general details see, e.g.,][]{jolliffe2002, jolliffe2016}.
Let $ \vec{I} (t) = (\vecChi(t),\vecPsi(t))$ be the 8-dimensional time series of the action variables evaluated along a numerical 
solution of the equations of motion. 
As in Sect.~\ref{sec:numerical_checks}, we apply the KZ low-pass filter with three iterations of the moving average and a cutoff 
frequency of 1 Myr$^{-1}$ to obtain the filtered time series $ \hat{\vec{I}} (t) $ \cite{Zurbenko2017, Mogavero2021}. In this way, 
the short-term quasi-periodic oscillations are mostly suppressed, which better reveals the chaotic diffusion over longer timescales. 
We finally define the mean-subtracted filtered action variables over the time interval $[t_0, t_0+T]$ as 
$\tilde{\vec{I}}(t) = \hat{\vec{I}}(t) - n^{-1} \sum_{i=0}^{n-1} \hat{\vec{I}}(t_0+i\Delta t)$, where the mean is estimated by 
discretization of the time series with a sampling step $\Delta t$ such that $T = (n-1)\Delta t$. The discretized time series over 
the given interval is stored in an $8 \times n$ matrix: 
\begin{equation} 
\label{eq:data}
\mat{D} = [\tilde{\vec{I}}(t_0),\ \tilde{\vec{I}}(t_0+\Delta t),\ \dots,\ \tilde{\vec{I}}(t_0+ (n-1)\Delta t)]. 
\end{equation}
The PCA of the data matrix $\mat{D}$ consists in a linear transformation $\mat{P} = \mat{A}^\mathrm{T} \mat{D}$, 
where $\mat{A}$ is an $8 \times 8$ orthogonal matrix (i.e. $\mat{A}^{-1} = \mat{A}^\mathrm{T}$) defined as follows. 
By writing $\mat{A} = [\vec{a}_1, \dots, \vec{a}_8]$, the column vectors $\vec{a}_i \in \mathbb{R}^8$ are chosen 
to be the normalized eigenvectors of the sample covariance matrix, in order of decreasing eigenvalues: 
$(n-1)^{-1} \mat{D}\mat{D}^\mathrm{T} = \mat{A} \vec{\Sigma} \mat{A}^\mathrm{T}$, 
where $\vec{\Sigma} = \text{diag} (\sigma_1,\dots,\sigma_8)$ and $\sigma_1 \geq \dots \geq \sigma_8$.
The PCs are defined as the new variables after the transformation, that is, $\text{PC}_i = \vec{a}_i \cdot \vec{I}$ 
with $i \in \{1,\dots,8\}$. The uncorrelatedness and the ordering of the PCs can be easily seen from the diagonal form 
of their sample covariance matrix, $(n-1)^{-1} \mat{P}\mat{P}^\mathrm{T} = \vec{\Sigma}$, from which it follows 
that the variance of $\text{PC}_i$ is $\sigma_i$. 

Among all the linear combinations in the action variables $\vec{I}$, the last PC, i.e., $\text{PC}_8$, has the smallest variance 
over the time interval $[t_0, t_0 + T]$ of a given orbital solution. The second last PC, i.e., $\text{PC}_7$, has the second smallest 
variance and is uncorrelated with $\text{PC}_8$, and so on. 
It follows that the linear subspace spanned by the last $k$ PCs is the $k$-dimensional subspace of minimum 
variance: the variance of the sample projection onto this subspace is the minimum among all the subspaces of the same dimension. 
These properties indicate that the last PCs provide a pertinent statistical definition of slow variables along numerically 
integrated solutions of a dynamical system. The linear structure of the PCA, in particular, seems adapted to 
quasi-integrable systems close to a quadratic Hamiltonian, like the ISS. In such a case, one may reasonably expect that the 
slow variables are, to a first approximation, linear combinations of the action variables. We remark that the mutual 
orthogonality allows us to associate a \textit{linear} d.o.f. to each PC. 

\paragraph*{Aggregated sample.}
Instead of considering a specific solution, it is also possible to take the same time interval from $m$ different 
solutions, and stack them together to form an aggregated sample: $ \mat{D}_\mathrm{agg} = [\mat{D}_1 , \mat{D}_2, \dots, \mat{D}_m]$, 
where $ \mat{D}_i$ is the data matrix of Eq.~\eqref{eq:data} for the $i$th solution. Since this work deals with a non-stationary 
dynamics, as the ISS ceaselessly diffuses in the phase space \cite{Hoang2022}, we always consider the same time interval for each 
of the $m$ solutions. The aggregated sample is useful in capturing globally the behavior of the dynamics, because it averages out 
temporary and rare episodes arising along specific solutions. 

\subsection{Principal components and quasi-integrals}
Both the QIs and the last PCs represent slow variables, but are established through two different methods. 
Equations~\eqref{eq:FLEs_QIs} and \eqref{eq:hierarchy_simpleH} claim that the QIs found semi-analytically in 
Sect.~\ref{sec:quasi_integrals} are among the slowest d.o.f. of the ISS dynamics. This naturally suggests to compare 
the three QIs with the three last PCs retrieved from numerically integrated orbital solutions. In this part, we 
first introduce the procedure that we implement to establish a consistent and systematic correspondence between 
QIs and PCs. We then present both a visual and a quantitative geometric comparison between them. 

\subsubsection{Tweaking the QIs} 
\label{subsec:PC_QI}
The three last components PC$_8$, PC$_7$, PC$_6$ are represented by the set of vectors 
$S_{\mathrm{PCs}}=\{\vec{a}_8, \vec{a}_7, \vec{a}_6\}$ 
belonging to $\mathbb{R}^8$. By construction, these PCs have a linear, hierarchical, and orthogonal structure. In other words: 
the PCs are linear combinations of the action variables $\vecI$; denoting by $\preceq$ the order of statistical variance, one has 
$\mathrm{PC}_8 \preceq \mathrm{PC}_7 \preceq \mathrm{PC}_6$; the unit vectors $(\vec{a}_i)_{i=6}^8$ are orthogonal to each other. 
On the other hand, the QIs of motion $\Cinc, \C_2, \mathcal{E}_{2n}$ do not possess these properties. Therefore, we adjust 
them in such a way to reproduce the same structure. 
\paragraph{Linearity.} 
While $\Cinc$ and $\C_2$ are linear functions of the action variables, $\mathcal{E}_{2n}$ is not when $n>1$. Nevertheless, 
as we explain in Sect.~\ref{sec:quasi_symmetries}, as far as one considers stable orbital solutions, the linear LL approximation 
$\mathcal{E}_2 = \vec{\gamma}_3 \cdot \vecI$ reproduces $\mathcal{E}_{2n}$ reasonably well. Therefore, we consider the 
three linear QIs of motion $\Cinc, \C_2, \mathcal{E}_2$, which are respectively represented by the set of $\mathbb{R}^8$-vectors 
$S_{\mathrm{QIs}}=\{\vec{\gamma}_1, \vec{\gamma}_2, \vec{\gamma}_3\}$. 
In this way, the 3-dimensional linear subspaces of the action space spanned by the sets $S_{\mathrm{QIs}}$ and $S_{\mathrm{PCs}}$ 
can be compared. 
\paragraph{Ordering.} We define a set of QIs that are ordered by statistical variance, as it is the case for the PCs. 
We follow two different approaches according to model $\simpleH{2n}$ in Eq.~\eqref{eq:simpleH} or $\Huv_{2n}$ in 
Eq.~\eqref{eq:hamiltonian} (clearly $n>1$). 
\begin{description}
\item[$\simpleH{2n}$] A strong hierarchy of statistical variances among the QIs emerges from the size of the leading symmetry-breaking 
resonances in Table~\ref{tab:table3} and from the orbital solutions in Figs.~\ref{fig:slowfast}, \ref{fig:PDF_QIs}, and \ref{fig:IQR_QIs}. 
One has $\mathcal{E}_{2n}^\bullet \prec \C_2 \prec \Cinc$. While $\mathcal{E}_{2n}^\bullet$ is an exact non-linear integral of motion, 
we expect that its linear truncation $\mathcal{E}_2^\bullet = \mathcal{E}_2$ varies more than $\C_2$ and $\Cinc$. 
Therefore, we consider the ordered set of QIs of motion $\{\C_2, \Cinc, \mathcal{E}_2\}$ represented by the ordered set of vectors 
$S'_{\mathrm{QIs}}=\{ \vec{\gamma}_2, \vec{\gamma}_1, \vec{\gamma}_3\}$. 
\item[$\Huv_{2n}$] Since the leading resonances affecting the QIs in Table~\ref{tab:table2} have comparable sizes, there is no clear 
order of statistical variances that can be inferred. We then implement a systematic approach that orders the QIs by simply inheriting the 
ordering of the PCs. More precisely, we define a set of ordered vectors $S_{\mathrm{QIs}}'$ through the projections of the three 
last PCs onto the linear subspace generated by the QIs: 
$S_{\mathrm{QIs}}'= \{ \mathrm{proj}_{S_{\mathrm{QIs}}} (\vec{a}_8),\mathrm{proj}_{S_{\mathrm{QIs}}} (\vec{a}_7), \mathrm{proj}_{S_{\mathrm{QIs}}} (\vec{a}_6) \}$\footnote{The projection of a vector $\vec{q}$ onto the subspace spanned by a set of vectors $S$ 
can be written in a vectorial form as $\mat{B} (\mat{B}^\mathrm{T} \mat{B})^{-1} \mat{B}^\mathrm{T} \vec{q}$, where the column space 
of the matrix $\mat{B}$ is the subspace spanned by the set $S$.}. 
As a result, the new set of QIs mirrors the hierarchical structure of the PCs. We stress that $S_{\mathrm{QI}}'$ spans the same 
subspace of $\mathbb{R}^8$ as $S_{\mathrm{QI}}$, since the ordered QIs are just linear combinations of the original ones. 
\end{description}

\paragraph{Orthogonality.} We apply the Gram-Schmidt process to the ordered set $S_{\mathrm{QIs}}'$ to obtain the orthonormal 
basis $S_{\mathrm{QIs}}''= \{ \vec{\alpha}_1, \vec{\alpha}_2, \vec{\alpha}_3 \}$. The set $S_{\mathrm{QIs}}''$ clearly spans the same 
subspace as $S_{\mathrm{QIs}}$. Moreover, the Gram-Schmidt process preserves the hierarchical structure, that is, the two $m$-dimensional 
subspaces spanned by the first $m \leq 3$ vectors of $S_{\mathrm{QIs}}'$ and $S_{\mathrm{QIs}}''$, respectively, are identical. 

In the end, we obtain a linear, ordered, and orthogonal set of \textit{modified} QIs of motion 
$\{ \mathrm{QI}_1, \mathrm{QI}_2, \mathrm{QI}_3 \}$, where $\mathrm{QI}_i = \vec{\alpha}_i \cdot \vec{I}$. 

\subsubsection{Visual comparison}
\label{sec:visual_comparison}
We now visually compare the vectors $\vec{\alpha}_{1,2,3}$ of the modified QIs with the corresponding vectors $\vec{a}_{8,7,6}$ 
of the last three PCs. We use the ensembles of 1080 numerically integrated orbital solutions of the models $\Hiss_4$ 
and $\Huv_4$ considered in Sects.~\ref{sec:numerical_checks} and \ref{sec:new_truncation}, respectively. The nominal solution 
of each set is denoted as sol.~\#1 from now on. For the model $\Hiss_4$, we also consider two other solutions: sol.~\#2 that 
represents a typical evolution among the 1080 solutions, and sol.~\#3 representing a rarer one. The particular choice of these 
two solutions is detailed in Sect.~\ref{sec:subspace_distance}. 

\begin{figure}
\includegraphics[width=\linewidth]{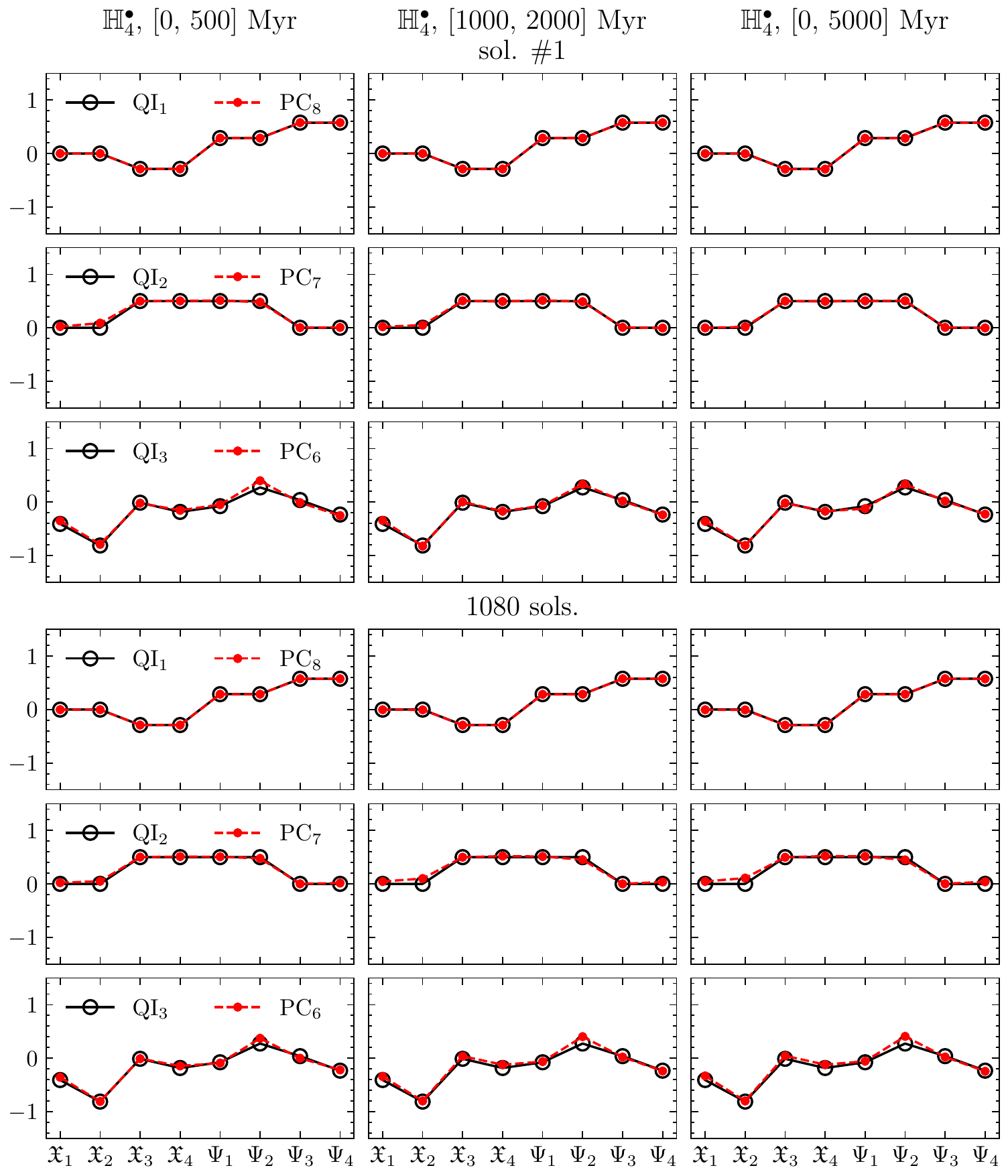}
\caption{\label{fig:PCQI_H4H} Vectors $\vec{\alpha}_{1,2,3}$ representing the three modified QIs (QI$_{1,2,3}$, black circles) 
compared to the corresponding vectors $\vec{a}_{8,7,6}$ of the three last PCs (PC$_{8,7,6}$, red dots), 
for the intervals [0, 500] Myr (left-hand column), [1000, 2000] Myr (middle column) and [0, 5000] Myr (right-hand column) of 
sol.~\#1 and of the aggregated sample of 1080 solutions of model $\simpleH{4}$. 
Here, $\mathrm{QI}_{1}$ is proportional to $\mathcal{C}_2$ and $\mathrm{QI}_{2}$ is proportional to $\mathcal{C}_2^{\perp}$; 
see Eq.~\eqref{eq:C4_C5}.} 
\end{figure}

\begin{figure}
\includegraphics[width=\linewidth]{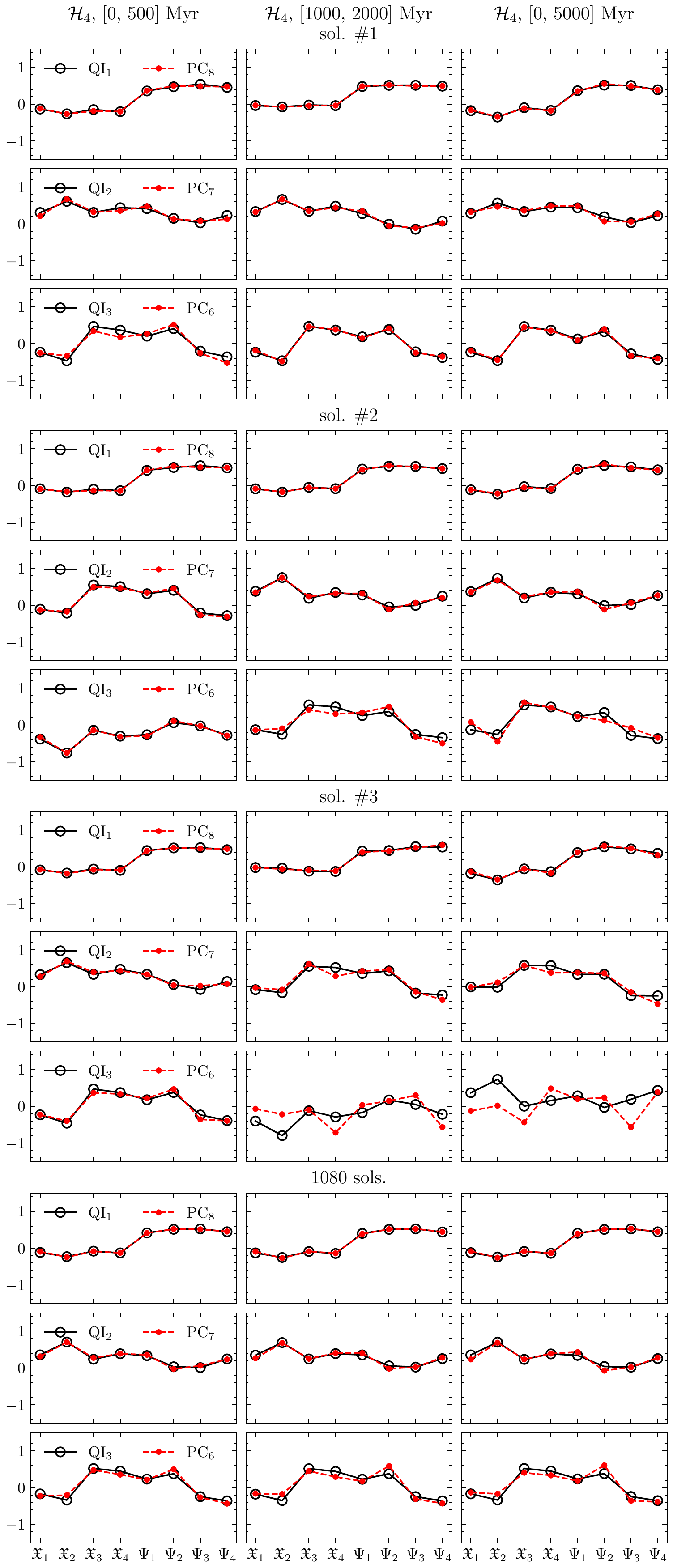}
\caption{\label{fig:PCQI_H4} Vectors $\vec{\alpha}_{1,2,3}$ representing the three modified QIs (QI$_{1,2,3}$, black circles) 
compared to the corresponding vectors $\vec{a}_{8,7,6}$ of the three last PCs (PC$_{8,7,6}$, red dots), 
for the intervals [0, 500] Myr (left-hand column), [1000, 2000] Myr (middle column) and [0, 5000] Myr (right-hand column) of 
sol.~\#1, sol.~\#2, and sol.~\#3 and of the aggregated sample of 1080 solutions of model $\Hiss_{4}$.} 
\end{figure}

\paragraph*{Hamiltonian $\simpleH{4}$.}
The modified QIs can be explicitly derived in this case and comprise interpretable physical quantities. 
One has $\mathrm{QI}_{1}$ proportional to $\mathcal{C}_2$ and $\mathrm{QI}_{2}$ proportional to $\mathcal{C}_2^{\perp}$. 
Moreover, $\mathrm{QI}_{3}$ is the component of $\mathcal{E}_2$ that is orthogonal to both $\mathcal{C}_2$ and $\mathcal{C}_2^{\perp}$. 
Figure~\ref{fig:PCQI_H4H} shows the comparison between the modified QIs and the corresponding PCs for three different time intervals 
along sol.~\#1 of $\simpleH{4}$ (see Fig.~\ref{fig:slowfast} bottom left for its time evolution). 
The agreement of the pairs $(\mathrm{QI}_{1},\mathrm{PC}_{8})$, $(\mathrm{QI}_{2},\mathrm{PC}_{7})$, and $(\mathrm{QI}_{3},\mathrm{PC}_{6})$ 
across different intervals is manifest and even impressive. 
One can appreciate that the ``slower" the PC, the more similar it is to its corresponding QI. 
The overlap between the modified QIs and the three last PCs means that the QIs of motion span the slowest 3-dimensional linear 
subspace of the action space. Therefore, to a linear approximation, they represent the three slowest d.o.f. of the $\simpleH{4}$ dynamics. 
The quasi-integral $\mathcal{C}_2$ represents the slowest linear d.o.f.: it coincides with the 
last principal component PC$_8$, which has the smallest variance among all the linear combinations of the action variables. 
$\mathcal{C}_{\mathrm{inc}}$ and $\mathcal{E}_2$ represent the second and the third slowest linear d.o.f., respectively: the component 
of $\mathcal{C}_{\mathrm{inc}}$ orthogonal to $\mathcal{C}_2$, i.e., $\C_2^\perp$, matches the second last principal component 
PC$_7$; the component of $\mathcal{E}_2$ orthogonal to the subspace generated by ($\mathcal{C}_2$, $\mathcal{C}_{\mathrm{inc}}$) 
matches the third last principal component PC$_6$. The strong hierarchical structure of the slow variables for the simplified 
dynamics $\simpleH{4}$ is clearly confirmed by the almost frozen basis vectors of the PCs.

\paragraph*{Hamiltonian $\Hiss_{4}$.}
In this case, the QIs of motion $\Cinc, \C_2, \mathcal{E}_2$ do not show a clear hierarchical structure in terms of statistical variance.  
Therefore, we consider the whole subspace spanned by the three QIs with respect to that spanned by the three last PCs. 
Since it is not easy to visually compare two 3-dimensional subspaces of $\mathbb{R}^8$, we compare their basis vectors instead. 
The basis $\vec{\alpha}_{1,2,3}$ of modified quasi-integrals QI$_{1,2,3}$ is computed according to the algorithm presented in 
Sect.~\ref{subsec:PC_QI}. 

\begin{figure}
\includegraphics[width=\linewidth]{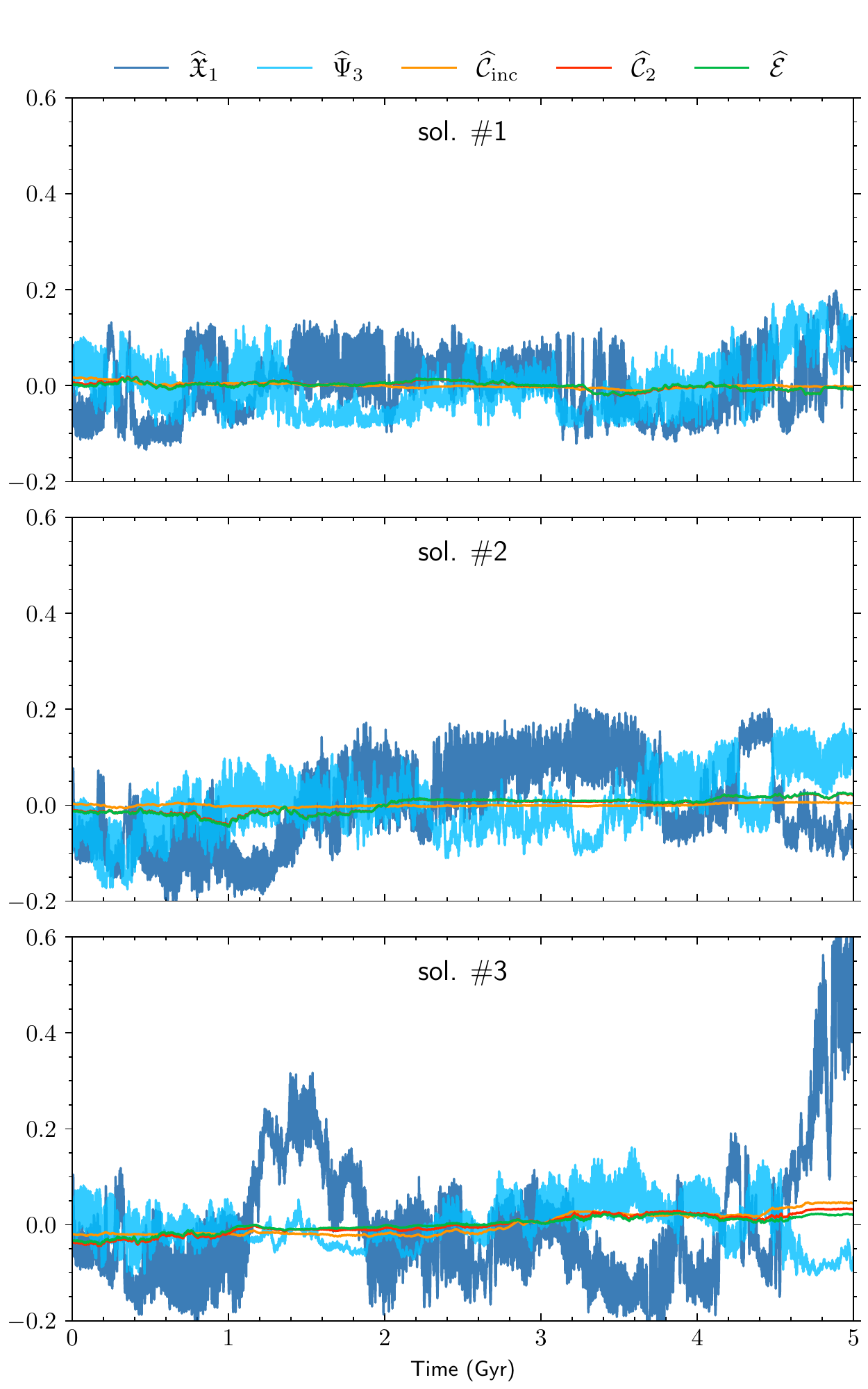}
\caption{\label{fig:QIs_H4} Time evolution over 5 Gyr of the 
dimensionless QIs of motions 
($\widehat{\mathcal{C}}_{\mathrm{inc}}, \widehat{\mathcal{C}}_2, \widehat{\mathcal{E}}$) and of two representatives of the 
dimensionless action variables ($\widehat{\Chi}_1, \widehat{\Psi}_3$) for three solutions of $\Hiss_4$, that is, sol.~\#1 (top), 
sol.~\#2 (middle), and sol.~\#3 (bottom). $\widehat{\mathcal{E}}$ stands for $\widehat{\mathcal{E}}_4$. 
The time series are low-pass filtered with a cutoff frequency of 1 Myr$^{-1}$ and the mean over 5 Gyr is subtracted.} 
\end{figure}

Figure~\ref{fig:PCQI_H4} presents the comparison between the modified QIs and the corresponding PCs across three different time intervals 
of three solutions of $\Hiss_4$ (see Fig.~\ref{fig:QIs_H4} for their time evolution). The first two, sols.~\#1 and \#2, show thorough 
agreement between the pairs of QIs and PCs across all intervals, which indicates close proximity between the two subspaces 
$V_{\mathrm{QIs}} = \Span(S_{\mathrm{QIs}})$ and $V_{\mathrm{PCs}} = \Span(S_{\mathrm{PCs}})$. One can appreciate that the directions of 
the basis vectors are quite stable. The last component PC$_8$, 
in particular, remains close to $\Cinc$. The slowest linear d.o.f. of $\Hiss_4$ can thus be deduced to be close to $\Cinc$, 
in line with the discussion in Sect.~\ref{sec:secondary_resonances}. Such a result shows how interesting physical insight 
can be gained through the PCA. 
Some changes in the basis vectors can arise, however, as for the first time interval of sol.~\#2. This may be expected from a dynamical 
point of view. Differently from $\simpleH{4}$, there is no pronounced separation between the slowest d.o.f. at the bottom of the Lyapunov 
spectrum in Fig.~\ref{fig:LCE4_1}: the marginal distributions of consecutive exponents can indeed touch or overlap each other. Therefore, 
the hierarchy of slow variables is not as frozen as in $\simpleH{4}$ and it can change along a given orbital solution. 

Solutions~\#1 and \#2 represent typical orbital evolutions. If the same time intervals of all the 1080 solutions are stacked together 
to form an aggregated sample on which the PCA is applied, the features mentioned above persist: the agreement between QIs and PCs, 
the stability of the basis vectors, and the similarity between PC$_8$ and $\mathcal{C}_{\mathrm{inc}}$ (see Fig.~\ref{fig:PCQI_H4}). 
Once again, the PCA confirms that the subspace spanned by the three QIs is overall close to the slowest 3-dimensional linear subspace 
of the action space. Therefore, to a linear approximation, they represent the three slowest d.o.f. of the $\Hiss_4$ dynamics. 
We remark that the slowness of the 3-dimensional subspace spanned by the QIs is a much stronger constraint than the observation that each 
QI is a slow variable. To give an example, let $Q = \widehat{\vec{q}} \cdot \vec{I}$ be a slow variable with unit vector $\widehat{\vec{q}}$. 
If $\vec{\epsilon}$ is an arbitrary small vector, i.e. $\Vert \vec{\epsilon} \Vert \ll 1 $, 
then $ Q' = (\widehat{\vec{q}}+\vec{\epsilon}) \cdot \vec{I}$ can also be considered as a slow variable, whereas the normalized difference 
of two quantities, $\widehat{\vec{\epsilon}} \cdot \vec{I}$, is generally not. Therefore, the linear subspace spanned by $Q$ and $Q^\prime$, that is, 
by $\widehat{\vec{q}}$ and $\widehat{\vec{\epsilon}}$, is not a slow 2-dimensional subspace. 

Solution~\#3 in Fig.~\ref{fig:PCQI_H4} represents an edge case (see Fig.~\ref{fig:QIs_H4} for its time evolution). 
Typically, the variances of the QIs are at least one order of magnitude smaller than those of the action variables, which allows a clear 
separation. Nevertheless, the distinction between the QIs and faster d.o.f. can be more difficult in two rare possibilities. 
Firstly, if the change in a QI accumulates continually in one direction, its variance can inflate over a long time interval. 
This is the case for the interval [0, 5] Gyr of sol.~\#3. Secondly, the variance of a variable that is typically fast can suddenly 
dwindle during a certain period of time, for example, $\Psi_3$ over the interval [1,~2] Gyr of sol.~\#3. In both cases, 
the slow subspace defined by the three last PCs can move away from the QI subspace due to the contamination by d.o.f. that 
are typically faster. This is reflected in the mismatch of QI$_3$ and PC$_6$ on the last two time intervals of sol.~\#3 in 
Fig.~\ref{fig:PCQI_H4}. We remark that PC$_{8,7}$ are still relatively close to QI$_{1,2}$, which indicates that the slowest 2-dimensional 
subspace spanned by PC$_{8,7}$ still resides inside the QI subspace. 
It should be stressed that this disagreement between QIs and PCs does not mean that the QIs are not slow variables in this case. 
The mismatch has a clear dynamical origin instead. The resonance tables of this work have been retrieved from a single, very long 
orbital solution, with the idea that its time statistics is representative of the ensemble statistics over a set of initially very 
close solutions \citep{Mogavero2022}. Therefore, the QIs derived from these tables characterize the dynamics in a global sense. 
The network of resonances can temporarily change in an appreciable way along specific solution, or be very particular along rare 
orbital solutions. In these cases, a mismatch between the last PCs and the present QIs may naturally arise. 
Moreover, the contamination of the QIs by d.o.f. that are typically faster may also be expected from the previously mentioned 
lack of a strong hierarchical structure of the slow variables. The Lyapunov spectrum in Fig.~\ref{fig:LCE4_1} shows that the 
marginal distributions of the exponents $\lambda_5$ and $\lambda_6$, for example, are not separate but overlap each other. 

\subsubsection{Distance between the subspaces of PCs and QIs}
\label{sec:subspace_distance}
The closeness of the two 3-dimensional linear subspaces $V_{\mathrm{PCs}}, V_{\mathrm{QIs}} \subset \mathbb{R}^8$ 
spanned by the sets of vectors $S_{\mathrm{PCs}}$ and $S_{\mathrm{QIs}}$, respectively, can be quantitatively measured in terms 
of a geometric distance. This can be formulated using the principal (canonical) angles \cite{jordan1875, van1996, ye2016}. 

Let $A$ and $B$ be two sets of $m \leq n$ independent vectors in $\mathbb{R}^n$. 
The principal vectors $(\vec{p}_k, \vec{q}_k)_{k=1}^m$ are defined recursively as solutions to the optimization problem: 
\begin{equation}
  \begin{alignedat}{2}
    & \text{maximize} & & \vec{p} \cdot \vec{q}\\
    & & \smash{\rule[-2.1\baselineskip]{0.6pt}{2.5\baselineskip}\ } \enspace & \vec{p} \in \Span(A), \enspace \vec{q} \in \Span(B), \\
    & \text{subject to} \enspace & & \Vert \vec{p} \Vert = 1, \enspace \Vert \vec{q} \Vert = 1, \\
    & & & \vec{p} \cdot \vec{p}_i = 0, \enspace \vec{q} \cdot \vec{q}_i = 0, \enspace i=1,\dots,k-1 , 
  \end{alignedat}
\end{equation}
for $k = 1, \dots, m$. The principal angles $0\leq \theta_1 \leq \dots \leq \theta_m \leq \pi/2$ between the two subspaces $\Span(A)$ 
and $\Span(B)$ are then defined by 
\begin{equation}
\cos \theta_k = \vec{p}_k \cdot \vec{q}_k, \quad k = 1, \dots, m .
\end{equation}
The principal angle $\theta_1$ is the smallest angle between all pairs of unit vectors in span($A$) and span($B$); the principal angle 
$\theta_2$ is the smallest angle between all pairs of unit vectors that are orthogonal to the first pair; and so on. Given the matrices defining 
the two subspaces, the principal angles can be computed from the singular value decomposition of their correlation matrix. The result is the 
canonical correlation matrix $\textrm{diag}(\cos \theta_1, \dots, \cos \theta_m)$. This cosine-based method is often ill-conditioned for small 
angles. In such case, a sine-based algorithm can be employed \cite{bjorck1973}. In this work, we use the combined technique detailed in 
Ref.~\cite{knyazev2002}. 

\begin{figure}
\includegraphics[width=\linewidth]{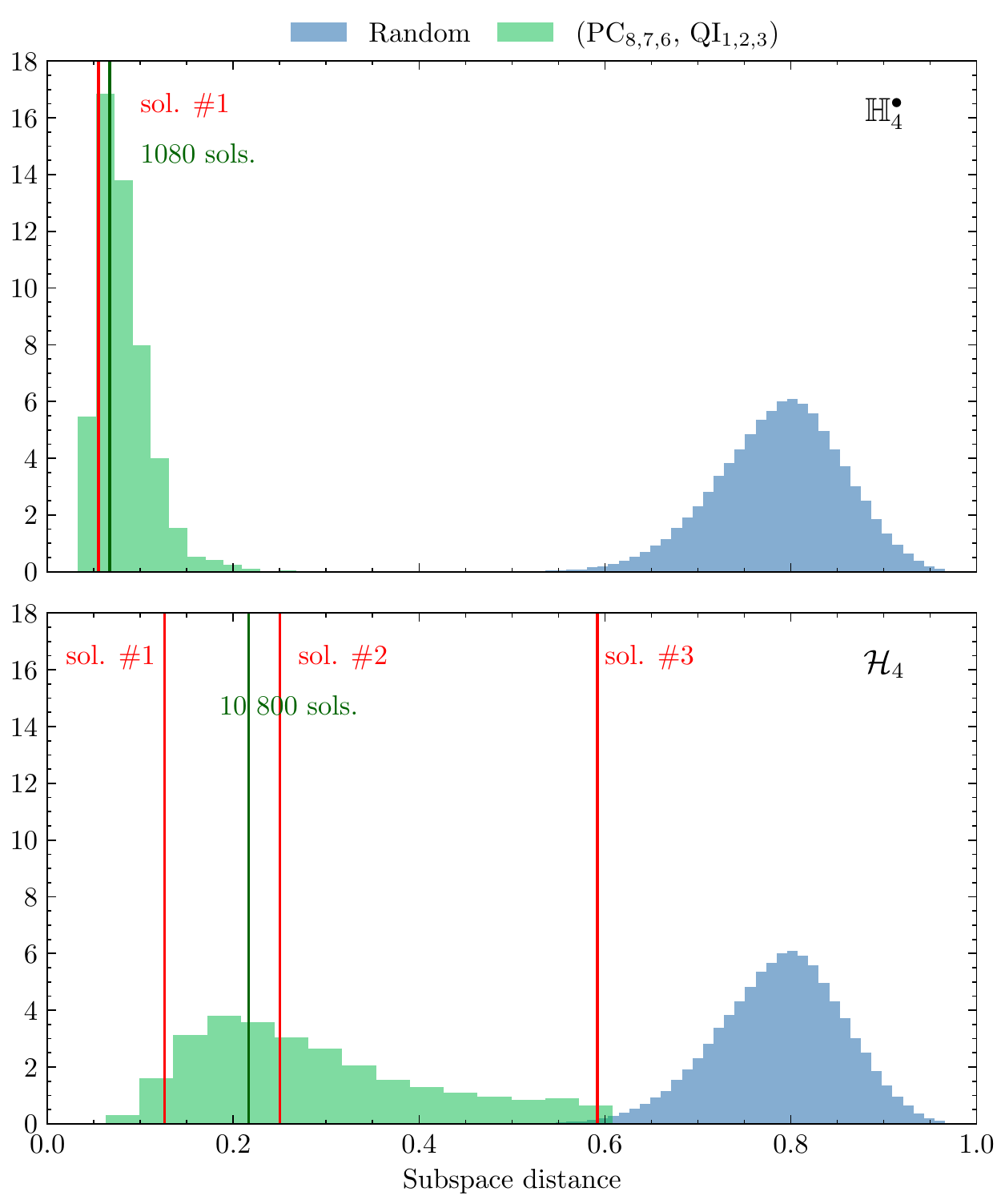}
\caption{\label{fig:PDF_metric} PDF of the distance between two random 3-dimensional linear subspaces 
of $\mathbb{R}^8$ (blue, $10^5$ draws) compared with the PDF of the distance between the two subspaces $V_{\mathrm{PCs}}$ 
(PC$_{8,7,6}$) and $V_{\mathrm{QIs}}$ (QI$_{1,2,3}$) arising from the time interval [0, 5] Gyr of 1080 solutions of $\simpleH{4}$ 
(top) and 10\,800 solutions of $\Hiss_4$ (bottom) (green). 
For each model, the subspace distance from the same time interval of representative solutions (vertical red lines) and of the aggregated sample 
of all the solutions (vertical dark green line) are indicated. The subspace distance is given by Eq.~\eqref{eq:distance}.} 
\end{figure}

Once the principal angles have been introduced, different metrics can be defined to measure the distance between two subspaces. In this work, 
we choose the normalized chordal distance \cite{ye2016}: 
\begin{equation} 
\label{eq:distance}
d(A, B) = \left(\frac{1}{m} \sum_{k=1}^m \sin^2 \theta_k  \right)^{1/2}.
\end{equation}
The distance is null if $A$ and $B$ are the same subspace and equal to $1$ when they are orthogonal. We use this metric to show 
that the subspace closeness suggested by Figs.~\ref{fig:PCQI_H4H} and \ref{fig:PCQI_H4} is indeed statistically significantly. 
More precisely, we provide evidence against the null hypothesis that the distribution of distances between $V_{\mathrm{PCs}}$ 
and $V_{\mathrm{PCs}}$, arising from the $\simpleH{4}$ and $\Hiss_4$ dynamics, coincides with that of randomly drawn 3-dimensional 
subspaces of $\mathbb{R}^8$. 
The PDF of the distance between two random 3-dimensional subspaces of $\mathbb{R}^8$ is shown in Fig.~\ref{fig:PDF_metric} in blue 
(such random subspaces can be easily generated by sampling sets of 3 vectors uniformly on the unit $7$-sphere \cite{Muller1959}). 
While the range of possible distances is [0,1], the distribution concentrates on the right-hand side of the interval, with a 
probability of approximately 99.3\% that the distance is larger than 0.6. In this regard, we remark that the notion of distance in 
high-dimensional spaces is very different from our intuition in a 3-dimensional world. If we draw randomly two vectors 
in a very high-dimensional space, it is extremely likely that they will be close to mutual orthogonality. 

The upper panel of Fig.~\ref{fig:PDF_metric} shows in green the PDF of the distance between $V_{\mathrm{PCs}}$ and $V_{\mathrm{QIs}}$ 
arising from the time interval [0, 5] Gyr of the 1080 orbital solutions of model $\simpleH{4}$. In the lower panel, we consider a larger 
ensemble of 10\,800 solutions of model $\Hiss_4$ spanning the same time interval \citep{Hoang2022}, and plot the corresponding PDF of 
the distance between $V_{\mathrm{PCs}}$ and $V_{\mathrm{QIs}}$. In both cases, the distance stemming from the aggregated sample of all 
the solutions is indicated by a vertical dark green line. We also report the distances from the specific solutions considered in 
Figs.~\ref{fig:PCQI_H4H} and \ref{fig:PCQI_H4} as vertical red lines. As the PDFs of both models peak at small distances, 
there results a strong evidence that the distribution of distances between the subspaces spanned by the PCs and the QIs is not that of 
random subspaces. In this sense, the closeness of the subspaces $V_{\mathrm{PCs}}$ and $V_{\mathrm{QIs}}$ is a statistically robust result. 
In the case of the simplified dynamics $\simpleH{4}$, the PDF peaks around a median of roughly 0.08 and has small variance. 
Switching to model $\Hiss_4$, the median increases to about 0.26 and the PDF is more spread out, with a long tail toward larger distances. 
The differences between the PDFs of the two models follow quite naturally the discussion in Sect.~\ref{sec:visual_comparison}: 
a quasi frozen hierarchy of the slowest variables for $\simpleH{4}$; a larger variance for $\Hiss_4$ related to contamination 
by d.o.f. that are typically faster and to variations of the resonant network with respect to the nominal solution of Gauss's dynamics 
which is used to infer the QIs. Solution~\#3 in Fig.~\ref{fig:PCQI_H4} represents in this sense an edge case of the distance 
distribution, while sol.~\#2 is a typical solution close to the PDF median. 

\section{Implications \\on long-term stability}
\label{sec:implications}
The existence of slow variables can have crucial implications on the stability of the ISS. The QIs of motion can 
effectively constraint in an adiabatic way the chaotic diffusion of the planet orbits over long timescales, forbidding in 
general a dynamical instability over a limited time span, e.g., several billions of years. 
Here we give compelling arguments for such a mechanism. 

\begin{figure}
\includegraphics[width=\linewidth]{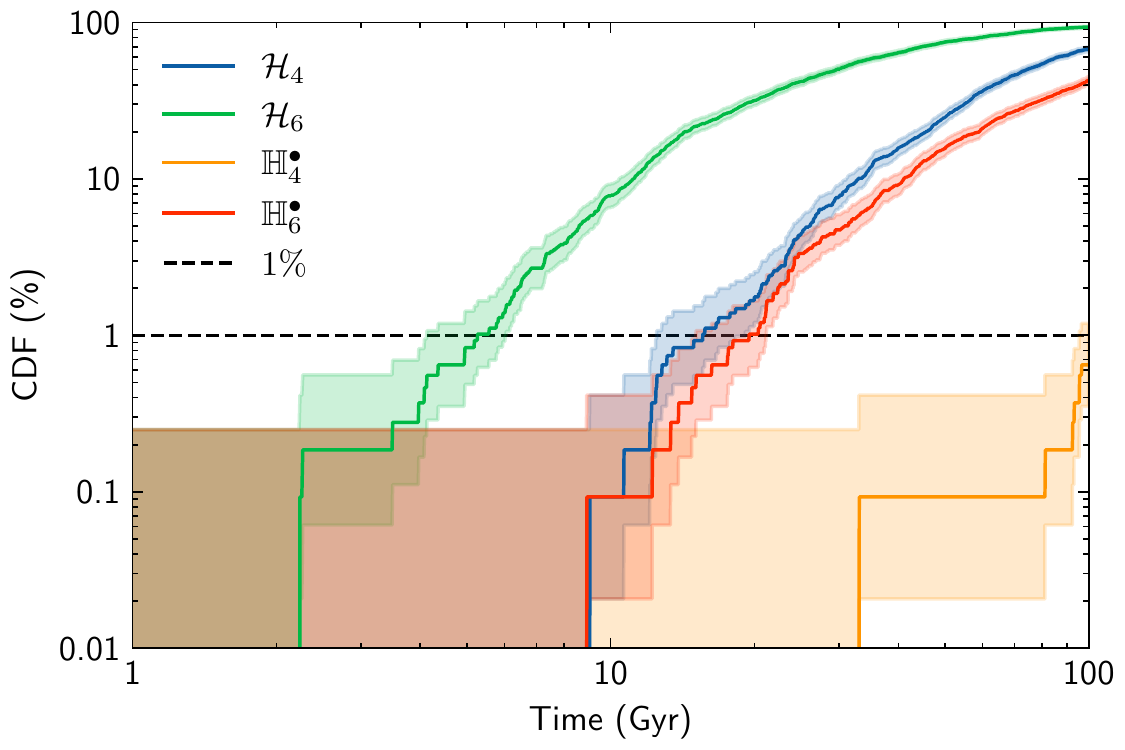}
\caption{\label{fig:CDFs} Cumulative distribution function of the first time that Mercury eccentricity reaches 
a value of 0.7, from 1080 orbital solutions of different models over 100 Gyr. The shaded regions represent the 
90\% piecewise confidence intervals from bootstrap.}
\end{figure}

Figure~\ref{fig:CDFs} shows the cumulative distribution function (CDF) of the first time that Mercury eccentricity reaches 
a value of 0.7, from the ensembles of 1080 orbital solutions of $\simpleH{4}$ and $\simpleH{6}$ introduced in 
Sect.~\ref{sec:new_truncation}. We recall that such a high eccentricity is a precursor of the dynamical instability 
(i.e., close encounters, collisions, or ejections of planets) of the ISS \citep{Laskar2009}. 
We also report the same CDF for the models $\Hiss_4$ and $\Hiss_6$, which we recently computed in Ref. \citep{Hoang2022}. 
One can appreciate that the time corresponding to a probability of instability of 1\% is greater than 100 Gyr for the $\simpleH{4}$ model, 
while it is about 15 Gyr for $\Hiss_4$. At degree 6, this time still increases from 5 Gyr for $\Hiss_6$ to about 20 Gyr in $\simpleH{6}$. 
The dynamics arising from $\simpleH{4}$ and $\simpleH{6}$ can be considered as stable in an astronomical sense. 
Recalling that the main difference between $\simpleH{2n}$ and $\Huv_{2n}$ relates to the smallest Lyapunov exponents 
(Fig.~\ref{fig:LCE4}), and this is accompanied by a much slower diffusion of the QIs for $\simpleH{2n}$ 
(Figs.~\ref{fig:slowfast}, \ref{fig:PDF_QIs}, and \ref{fig:IQR_QIs}), Fig.~\ref{fig:CDFs} indicates that the dynamical 
half-life of the ISS is linked to the speed of diffusion of these slow quantities in a critical way. We stress that the 
slower diffusion toward the dynamical instability in the $\simpleH{2n}$ model derives from neglecting the external forcing mainly 
exerted by Saturn, Uranus, and Neptune. 

We also observe that, to a linear approximation, the knowledge of $\Cinc$ and $\mathcal{E}_2$ allows us to bound the variations 
of the action variables $\vec{\Chi},\vec{\Psi}$. 
Recalling that the actions are positive quantities, from Eq.~\eqref{eq:Cinc} one sees that fixing a value of $\Cinc$ 
puts an upper bound to the variations of the inclination actions $\vec{\Psi}$. As a consequence, at degree 2 in eccentricities 
and inclinations, fixing a value of the QI 
\begin{equation}
\label{eq:QI_energy_LL_v2}
\mathcal{E}_2 = \vec{\gamma}_3 \cdot \vecI = 
\superscript{\vec{\gamma}}{ecc}_3 \cdot \vec{\Chi} + \superscript{\vec{\gamma}}{inc}_3 \cdot \vec{\Psi} ,
\end{equation}
with $\vec{\gamma}_3 = (\superscript{\vec{\gamma}}{ecc}_3, \superscript{\vec{\gamma}}{inc}_3)$, also bounds the upper variations 
of the eccentricity actions $\vec{\Chi}$, since the components of $\superscript{\vec{\gamma}}{ecc}_3$ have all the same sign, as those of 
$\superscript{\vec{\gamma}}{inc}_3$ (see Appendix~\ref{sec:gamma_vectors}). 
This is an important point, as we state in Sect.~\ref{sec:intro} that the lack of any bound on the chaotic variations 
of the planet orbits is one of the reasons that complicate the understanding of their long-term stability. We remark that the secular 
planetary phase space can be bound by fixing the value of the total AMD, that is, $\Cecc + \Cinc$ \citep{Laskar1997}. 
A statistical study of the density of states that are \textit{a priori} accessible can then be realized \citep{Mogavero2017}. 
It is not, however, fully satisfying to consider a fixed value of total AMD of the ISS, as we show that $\Cecc$ is changed by some of 
the leading resonances of the Hamiltonian, as a result of the eccentricity forcing mainly exerted by Jupiter through the mode $g_5$. 
Moreover, the destabilization of the ISS consists indeed in a large transfer of eccentricity AMD, $\Cecc$, from the outer system to the inner 
planets through the resonance $g_1-g_5$ \citep{Laskar2008,Batygin2008,Laskar2009,Boue2012}. 
It should be noted that $\Cecc$ can still be consider as a slow quantity 
with respect to an arbitrary function of the action variables, as it is only changed by the subset of the leading resonances involving 
the external mode $g_5$. This slowness has indeed been observed on stable orbital solutions of the Solar System \citep{Laskar1997} 
and supports the statistical hypothesis in Ref.~\citep{Mogavero2017} that allows one to obtain a very reasonable first guess of the long-term 
PDFs of the eccentricities and inclinations of the inner planets. 

The emerging picture explains the statistical stability of the ISS over billions of years in a physically intuitive way. 
The chaotic behavior of the planet orbits arises from the interaction of a number of leading resonant harmonics of the Hamiltonian, 
which determine the Lyapunov time. The strongest resonances are characterized by some exact symmetries, which are only broken by weak 
resonant interactions. These quasi-symmetries naturally give birth to QIs of motion, quantities that diffuse much more slowly 
than the LL action variables, constraining the variations of the orbits. The long dynamical half-life of the ISS is connected 
to the speed of this diffusion, which eventually drives the system to the instability. 
It should be stressed that, besides the speed of diffusion, the lifetime of the inner orbits also depends on the initial distance of 
the system from the instability boundary defined by the resonance $g_1-g_5$. This geometric aspect includes the stabilizing 
role of general relativity \citep{Laskar2008,Laskar2009}, which moves the system away from the instability boundary by 0.43\arcsecyr{}, 
and the destabilizing effect of terms of degree 6 in eccentricities and inclinations of the planets \citep{Hoang2022}. 

\section{Discussion}
\label{sec:discussion}
This work introduces a framework that naturally justifies the statistical stability shown by the ISS over a timescale 
comparable to its age. 
Considering a forced secular model of the inner planet orbits, the computation of the Lyapunov spectrum indicates the 
existence of very different dynamical timescales. Using the computer algebra system TRIP, we systematically analyze the Fourier 
harmonics of the Hamiltonian that become resonant along a numerically integrated orbital solution spanning 5 Gyr. We uncover three 
symmetries that characterize the strongest resonances and that are broken by weak resonant interactions. These quasi-symmetries generate 
three QIs of motion that represent slow variables of the secular dynamics. The size of the leading symmetry-breaking resonances 
suggests that the QIs are related to the smallest Lyapunov exponents. The claim that the QIs are among the slowest d.o.f. of the 
dynamics constitutes the central point of this work. On the one hand, it is supported by the analysis of the underlying 
Hamiltonian $\simpleH{2n}$, in which one neglects the forcing mainly exerted by Saturn, Uranus, and Neptune and, as a consequence, 
the diffusion of the QIs is greatly reduced. On the other hand, the geometric framework established by the PCA of the orbital 
solutions independently confirms that the QIs are statistically the slowest linear variables of the dynamics. We give strong evidence 
that the QIs of motion play a critical role in the statistical stability of the ISS over the Solar System lifetime, by adiabatically 
constraining the long-term chaotic diffusion of the orbits. 

\subsection{Inner Solar System among \\classical quasi-integrable systems}
\label{sec:FPUT}
It is valuable to contextualize the dynamics of the ISS in the class of classical quasi-integrable systems. 
A comparison with the Fermi-Pasta-Ulam-Tsingou problem, in particular, deserves to be made. 
This concerns the dynamics of a one-dimensional chain of identical masses coupled by nonlinear springs. 
For weak nonlinearity, the normal modes of oscillation remain far from the energy equipartition expected from 
statistical mechanics for a very long time \citep{Fermi1955}. 
One way to explain the lack of energy equipartition reported by Fermi and collaborators is through the closeness of the FPUT problem 
to the integrable Toda dynamics \citep{Henon1974,Flaschka1974,Manakov1974}. This translates in a very slow thermalization of the action 
variables of the Toda problem and of the corresponding integrals of motion along the FPUT flow 
\citep{Manakov1974,Ferguson1982,Benettin2013,Goldfriend2019,Chris2019,Grava2020}. 
In the framework of the present study, the very long dynamical half-life of the ISS is also likely to be the result 
of the slow diffusion of some dynamical quantities, the QIs of motion. We find, in particular, an underlying Hamiltonian $\simpleH{2n}$ 
for which this diffusion is greatly reduced, as a consequence of neglecting the forcing mainly exerted by Saturn, Uranus, and Neptune. This 
results in a dynamics that can be considered as stable in an astronomical sense. We stress that, differently from the FPUT 
problem, $\simpleH{2n}$ is not integrable as the Toda Hamiltonian. It is indeed chaotic and shares with the original 
Hamiltonian $\Huv_{2n}$ the leading Lyapunov exponents. The QIs that we find in this work are only a small number 
of functions of the action-angle variables of the integrable LL dynamics, and are related to the smallest Lyapunov exponents 
of the dynamics. Our study suggests that in the FPUT problem the very slow thermalization occurring beyond the Lyapunov time 
might be understood in terms of combinations of the Toda integrals of motion diffusing over very different timescales. 

The long-term diffusion in chaotic quasi-integrable systems should be generally characterized by a broad range of timescales 
that results from the progressive, hierarchical breaking of the symmetries of the underlying integrable problem by resonant 
interactions \citep{Ford1961,Onorato2015,Pistone2019}. A hierarchy of Lyapunov exponents spanning several orders of magnitude, 
in particular, should be common among this class of systems \citep[e.g.,][]{Malishava2022}. 

\subsection{Methods}
\label{sec:methods}
The long-term dynamics of the ISS is described by a moderate but not small number of d.o.f., which places it far from 
the typical application fields of celestial mechanics and statistical physics. The first discipline often studies dynamical models 
with very few degrees of freedom, while the second one deals with the limit of a very large number of bodies. Chaos also requires a 
statistical description of the inner planet orbits. But the lack of a statistical equilibrium, resulting from a slow but ceaseless 
diffusion of the system, places the ISS outside the standard framework of ergodic theory. 
The kind of approach we develop in this work is heavily based on computer algebra, in terms of systematic series expansion of the 
Hamiltonian, manipulation of the truncated equations of motion, extraction of given Fourier harmonics, retrieval of polynomial roots, 
etc. \citep{Mogavero2021,Mogavero2022}. This allows us to introduce QIs of motion in a 16-dimensional dynamics by analyzing how 
action-space symmetries are progressively broken by resonant interactions. Our effective method based on the time statistics of 
resonances arising along a single, very long numerical integration is alternative to formal approaches that define QIs via series 
expansions \citep[e.g.][]{Contopoulos1960,Kruskal1962}. The practical usefulness of these formal expansions for a dynamics that covers 
an intricate, high-dimensional network of resonances seems indeed doubtful. Through the retrieval of the half-widths of the 
symmetry-breaking resonances, computer algebra also permits us to extend the correspondence between the Lyapunov spectrum and the 
spectrum of resonances well beyond the standard relation linking the Lyapunov time to the strongest resonances\footnote{In this regard, 
it should be noted that a relation between QIs and Lyapunov exponents has already been highlighted in simple systems 
\citep{Contopoulos1978,Benettin1980}}. 

In the context of dynamical systems with a number of d.o.f. that is not small, this work also considers an approach based on PCA. 
The role of this statistical technique can be twofold. We use PCA as an independent test to systematically validate the slowness 
of the QIs. While being introduced semi-analytically as dynamical quantities that are not affected by the leading resonances, they can 
indeed be related to the last PCs. By extension, the first PCs should probe the directions of the main resonances. 
This leads to a second potential application of the PCA, which should offer a way to retrieve the principal resonant structure of 
a dynamical system. In this sense, PCA represents a tool to systematically probe numerical integrations of a complex dynamics and 
distill important hidden insights. 
We emphasize that PCA is the most basic linear technique of dimensionality reduction and belongs to the more general class of the 
unsupervised learning algorithms. There are more sophisticated methods of feature extraction that can be more robust 
\citep[e.g.][]{candes2011, markopoulos2014} and can incorporate nonlinearity \citep{lee2007}. These methods are often less 
intuitive to understand, less straightforward to apply, and harder to interpret than PCA. Yet, they might be more effective and 
worth pursuing for future works. 

With long-term numerical integration and a computer algebra system at one's disposal, the entire strategy we develop in this work 
can in principle be applied to other planetary systems and quasi-integrable Hamiltonian dynamics with a moderate number of d.o.f. 

\begin{acknowledgments}
The authors thank M. Gastineau for his assistance with \TRIP{}. 
F.M. is supported by a grant of the French Agence Nationale de la Recherche (AstroMeso ANR-19-CE31-0002-01). 
N.H.H. is supported by a Ph.D. scholarship of the Capital Fund Management (CFM) Foundation for Research.
This project has been supported by the European Research Council (ERC) 
under the European Union’s Horizon 2020 research and innovation program (Advanced Grant AstroGeo-885250). 
This work was granted access to the HPC resources of MesoPSL financed by the Region 
Île-de-France and the project Equip@Meso (reference ANR-10-EQPX-29-01) of the 
program Investissements d’Avenir supervised by the Agence Nationale pour la Recherche. 
\end{acknowledgments}

\appendix

\section{Lyapunov spectrum}

\begin{figure*}
\subfloat[\label{fig:LCE4_test1} $\Hiss_4$: Single orbital solution]{\includegraphics[width=0.99\columnwidth]{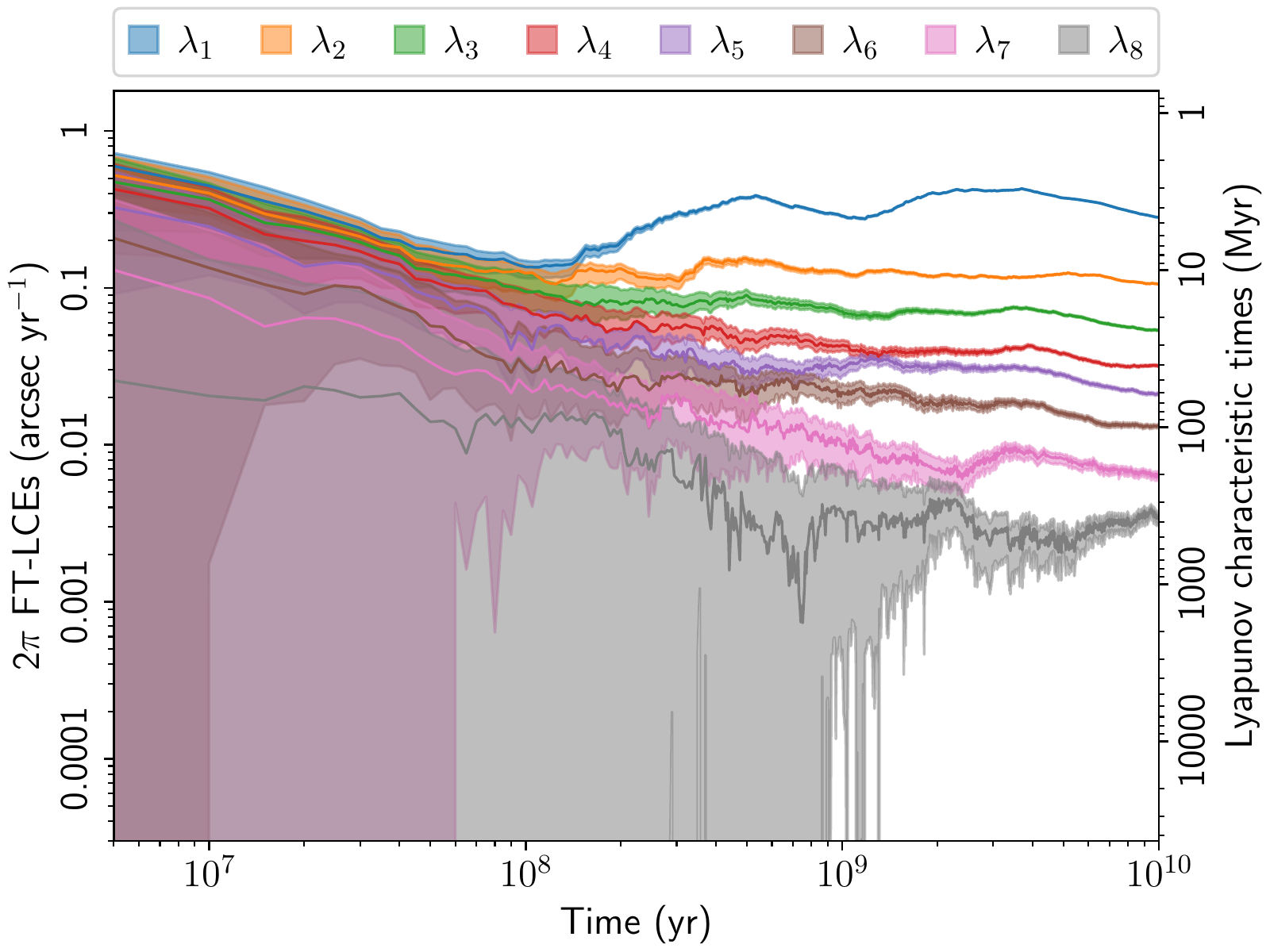}} 
\hphantom{ciao}
\subfloat[\label{fig:LCE4_test2} $\Hiss_4$: Relative numerical errors]{\includegraphics[width=0.99\columnwidth]{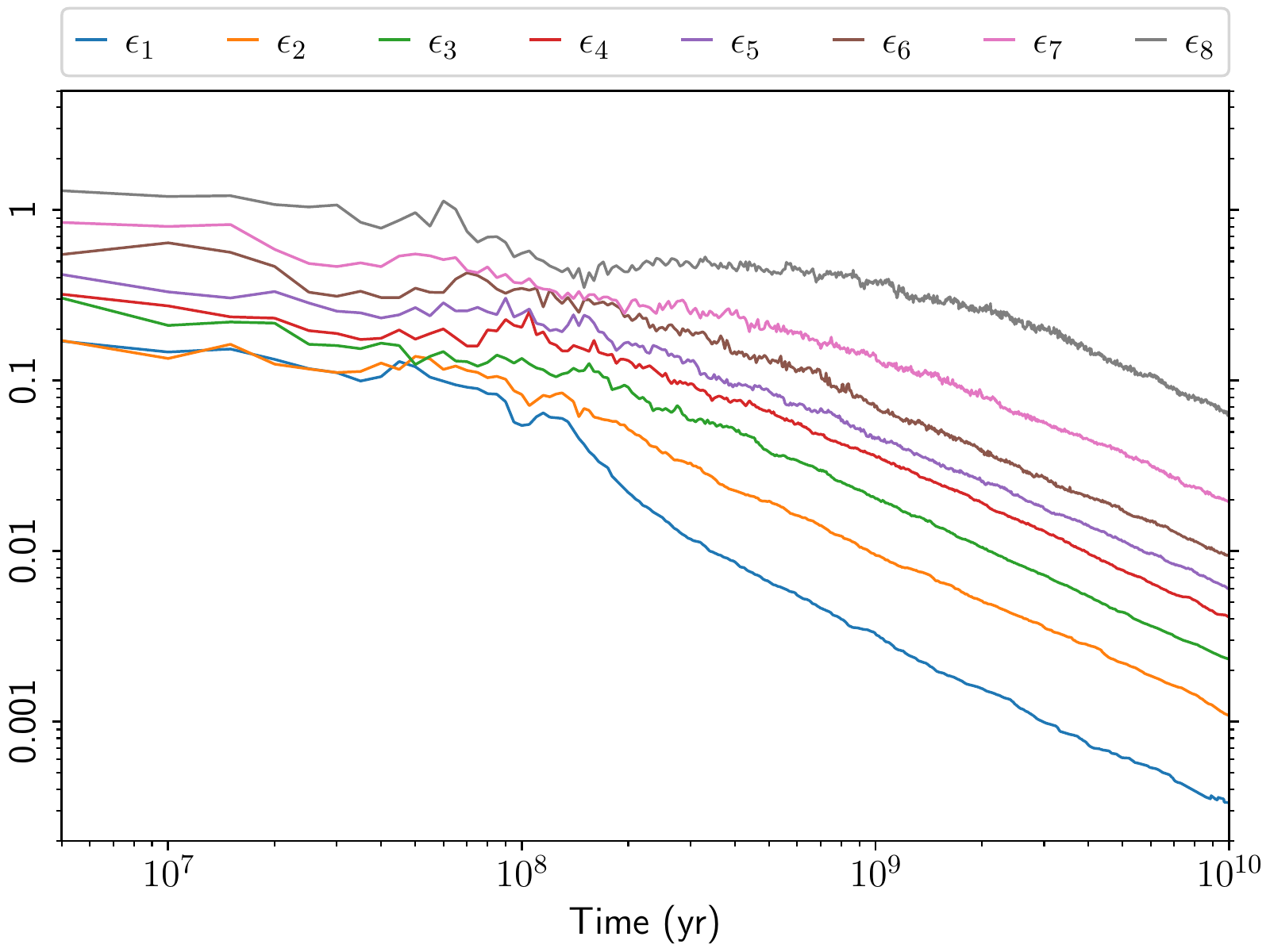}}
\caption{(a) Positive FT-LCEs of Hamiltonian $\Hiss_4$ and corresponding characteristic timescales 
for a single initial condition and an ensemble of 150 random sets of initial tangent vectors. The bands 
represent the [5th, 95th] percentile range of the marginal PDFs. The lines denote the distribution medians. 
(b) Medians of the relative numerical errors $\epsilon_i$ on the FT-LCEs $\lambda_i$, 
as defined in Eq.~\eqref{eq:LCE_relative_errors}, for the ensemble of 150 orbital solution of Fig.~\ref{fig:LCE4_1}.} 
\label{fig:LCE4_tests}
\end{figure*}

\paragraph*{Convergence.}
\label{sec:LCE_convergence}
We perform two tests to address the convergence of our implementation of the \citet{Benettin1980} method. 
We first compute the FT-LCEs for a single initial condition of $\Hiss_4$ and an ensemble of 150 different random sets 
of initial tangent vectors. Figure~\ref{fig:LCE4_test1} shows the [5th, 95th] percentile range of the resulting marginal 
distributions of the positive FT-LCEs over a time span of 10 Gyr. The distributions shrink with increasing time, 
eventually collapsing on single time-dependent 
values. In this asymptotic regime, the \citet{Benettin1980} algorithm loses memory of the initial tangent vectors and 
purely retrieves the FT-LCEs as defined in Eq.~\eqref{eq:FT_LCEs}. 
Therefore, Fig.~\ref{fig:LCE4_1} shows asymptotically the dependence of the FT-LCEs on the initial condition $\vec{z}_0$ 
and represents their statistical distribution over the phase-space domain explored by the dynamics in a non-ergodic way. 
The convergence of the computation is clearly slower for smaller exponents, but a comparison with Fig.~\ref{fig:LCE4_1} 
indicates that, even in the case of $\lambda_8$, the numerical uncertainty on the FT-LCEs of each orbital solution 
at 10 Gyr is negligible with respect to the width of their ensemble distributions. 

To quantitatively estimate the numerical precision on the computed FT-LCEs, we exploit the symmetry of the spectrum stated 
in Eq.~\eqref{eq:LCE_symmetry}. For a single orbital solution, the relative numerical error on each exponent $\lambda_i$ can 
be estimated as 
\begin{equation}
\label{eq:LCE_relative_errors}
\epsilon_i = \left| \frac{\Delta \lambda_i}{\lambda_i} \right| .
\end{equation}
We plot in Fig.~\ref{fig:LCE4_test2} the medians of $\epsilon_i$ for the ensemble of 150 orbital solutions of Fig.~\ref{fig:LCE4_1}. 
The relative errors decrease asymptotically with time, as expected. Even in the case of the smallest exponent, $\lambda_8$, 
the median error is less than 10\% at 10 Gyr. 

\begin{figure}
\centering
\includegraphics[width=\columnwidth]{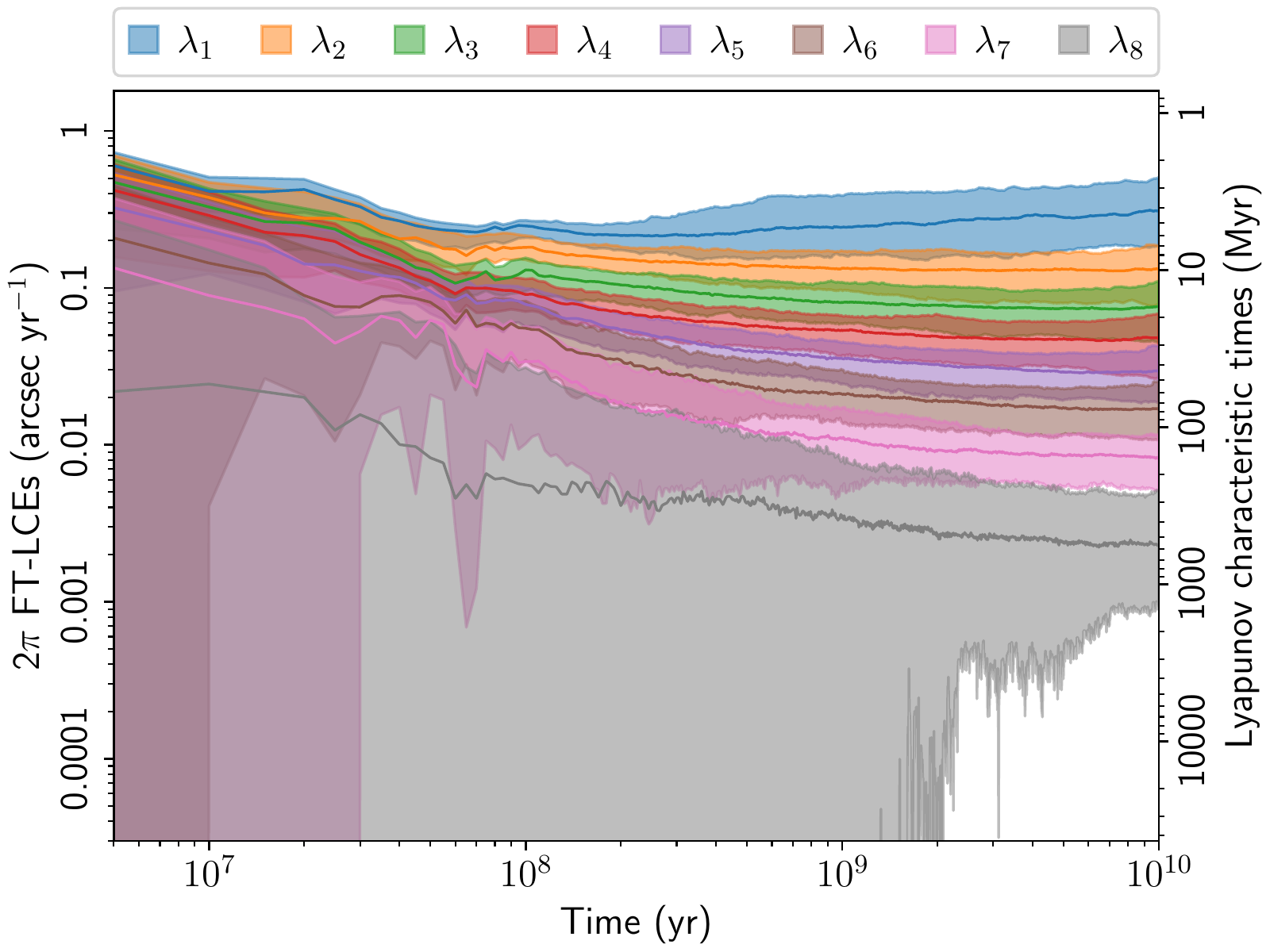}
\caption{Positive FT-LCEs $\lambda_i$ of Hamiltonian $\Hiss_6$ 
and corresponding characteristic timescales $\lambda_i^{-1}$. 
The bands represent the [5th, 95th] percentile range of the marginal PDFs estimated 
from an ensemble of 150 stable orbital solutions with very close initial conditions. 
The lines denote the distribution medians.} 
\label{fig:LCE6}
\end{figure}

\paragraph*{Hamiltonian $\Hiss_6$.}
\label{sec:LCE_6}
We compute for comparison the FT-LCEs of the forced ISS truncated at degree 6 in eccentricities and inclinations, that is, $\Hiss_6$. 
We consider 150 stable orbital solutions with initial conditions very close to the nominal values of Gauss's dynamics 
and random sets of initial tangent vectors, as we do for the truncation at degree 4. Figure~\ref{fig:LCE6} shows the [5th, 95th] percentile 
range of the marginal PDF of each FT-LCE estimated from the ensemble of solutions. Apart from being somewhat 
larger, the asymptotic distributions of the exponents are very similar to those of $\Hiss_4$ shown in Fig.~\ref{fig:LCE4_1}. 

\section{Vectors $\bgamma_1$, $\bgamma_2$, $\bgamma_3$}
\label{sec:gamma_vectors}
We report here the explicit expressions of the vectors $(\vec{\gamma}_i)_{i=1}^3$. 
We first give the components of the vector $\LL{\vec{\omega}}$ of the fundamental precession frequencies of the inner orbits 
in the forced Laplace-Lagrange dynamics (including the leading correction of general relativity) \citep{Mogavero2021}: 
\begin{equation}
\label{eq:gammaLL}
\begin{aligned}
\LL{\vec{\omega}} &= (\LL{\vec{g}}, \LL{\vec{s}}) \approx \\ 
&(5.87, 7.46, 17.4, 18.1, -5.21, -6.59, -18.8, -17.7),
\end{aligned}
\end{equation}
in units of \arcsecyrtext{} (see \citep{Brumberg1973,Bretagnon1974,Laskar1985} for comparison with the frequencies 
of the Laplace-Lagrange dynamics of the entire Solar System). One then has 
\begin{equation}
\label{eq:gamma_vectors}
\begin{aligned}
\vec{\gamma}_1 &= (\vec{0}_4, \vec{1}_4) = (0,0,0,0,1,1,1,1), \\
\vec{\gamma}_2 &= (0,0,-1,-1,1,1,2,2), \\
\vec{\gamma}_3 &= -\LL{\vec{\omega}} + g_5 \vec{1}_8 \approx \\ 
&(-1.61, -3.20, -13.2, -13.9, 9.47, 10.8, 23.0, 22.0), 
\end{aligned}
\end{equation}
with the components of $\vec{\gamma}_3$ in units of \arcsecyrtext{}. We recall that $g_5 \approx 4.257$\arcsecyr{} 
is a constant in the forced model of the ISS. The corresponding unit vectors $(\widehat{\vec{\gamma}}_i)_{i=1}^3$ are given by 
\begin{equation}
\label{eq:gamma_unit_vectors}
\begin{aligned}
\widehat{\vec{\gamma}}_1 &= (0,0,0,0,1,1,1,1) / 2, \\
\widehat{\vec{\gamma}}_2 &= (0,0,-1,-1,1,1,2,2) / 2 \sqrt{3}, \\
\widehat{\vec{\gamma}}_3 &\approx (-0.04, -0.08, -0.33, -0.35, 0.24, 0.27, 0.58, 0.55). 
\end{aligned}
\end{equation}
Since $1/2\sqrt{3} \approx 0.289$, the components of $\widehat{\vec{\gamma}}_3$ are only a few percent away 
from those of $\widehat{\vec{\gamma}}_2$. Therefore, along stable orbital solutions with typical bounded 
variations of the Mercury-dominated action variable $\Chi_1$, the two quantities $\C_2$ and $\mathcal{E}_{2n}$ 
exhibit very similar time evolutions. This is not the case anymore when Mercury orbit reaches high eccentricities. 

\section{Ensemble distributions of the quasi-integrals over time}
\label{sec:ensemble_stats_QIs}
\begin{figure*}
\includegraphics[width=\textwidth]{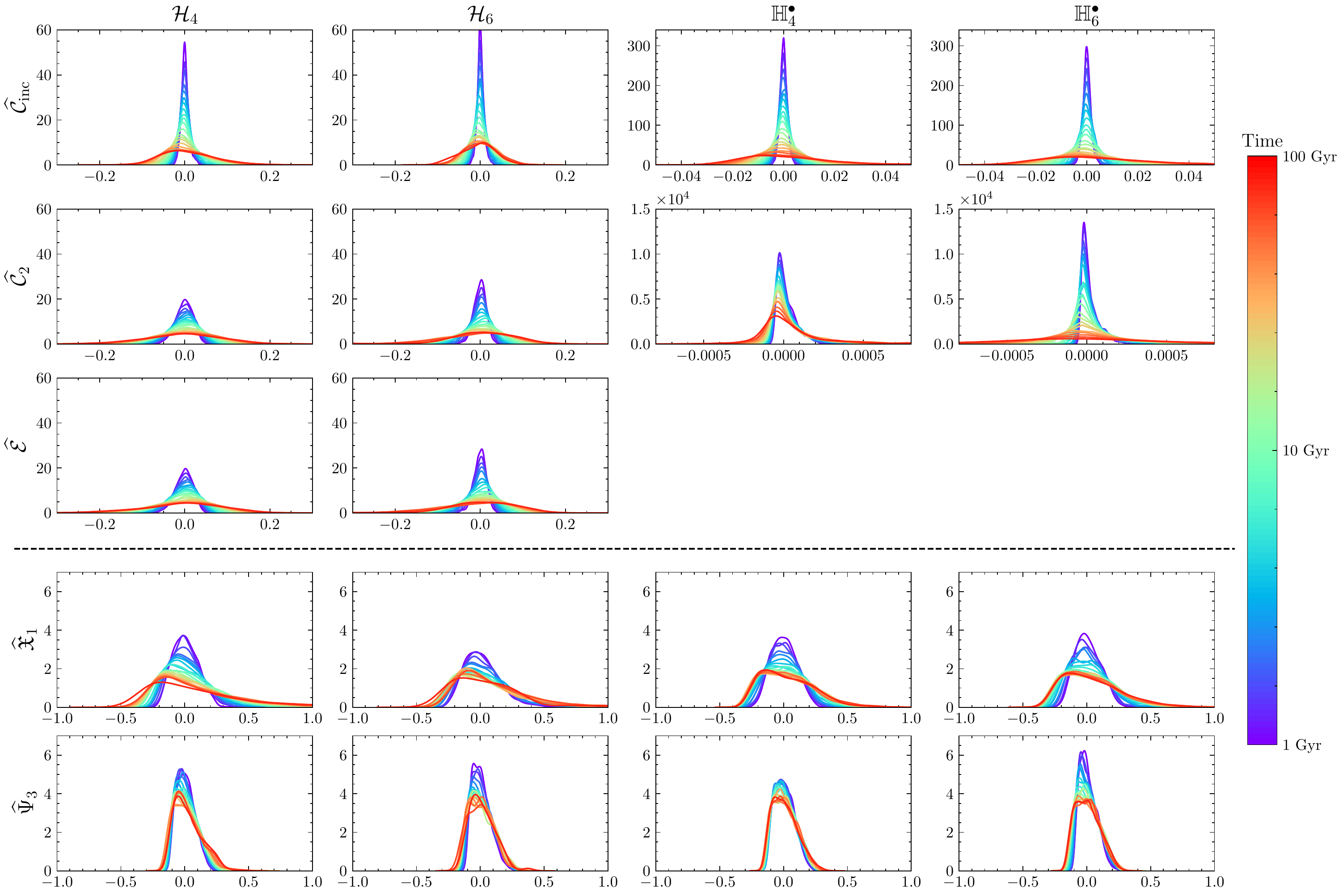}
\caption{\label{fig:PDF_QIs} 
Time evolution over 100 Gyr of the PDF of the signed deviation from the mean of the low-pass filtered dimensionless QIs and 
dimensionless actions $\Chi_1,\Psi_3$. 
Estimation from an ensemble of 1080 numerical orbital solutions for different models 
($\Hiss_4$, $\Hiss_6$, $\simpleH{4}$, and $\simpleH{6}$). 
\textit{First row}: $\Cinchat$. \textit{Second row}: $\widehat{\mathcal{C}}_2$. 
\textit{Third row}: $\widehat{\mathcal{E}}_4$ ($\Hiss_4$) and $\widehat{\mathcal{E}}_6$ ($\Hiss_6$). 
\textit{Fourth row}: $\widehat{\Chi}_1$. \textit{Fifth row}: $\widehat{\Psi}_3$. 
The time of each curve is color coded. At each time, the estimation only takes into account stable solutions, that are those with 
a running maximum of Mercury eccentricity smaller than 0.7. The quantity $\widehat{\mathcal{E}}_{2n}^\bullet$ is an exact 
integral of motion for the model $\simpleH{2n}$ and its PDF has null dispersion.}
\end{figure*}

\begin{figure*}
\includegraphics[width=\textwidth]{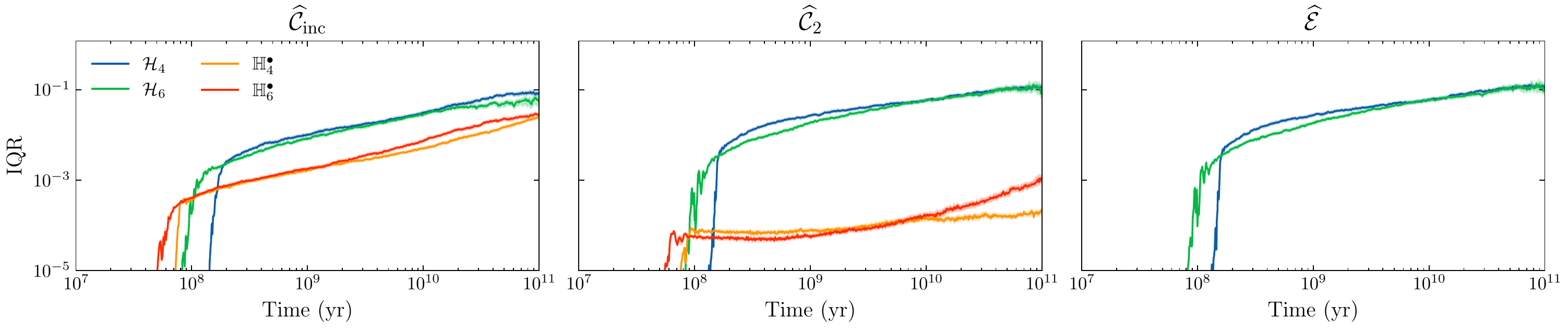}
\caption{\label{fig:IQR_QIs}
Time evolution of the interquartile range (IQR) of the ensemble PDFs of the QIs shown in Fig.~\ref{fig:PDF_QIs}. 
\textit{Left}: $\Cinchat$. 
\textit{Middle}: $\widehat{\mathcal{C}}_2$. 
\textit{Right}: $\widehat{\mathcal{E}}_4$ ($\Hiss_4$) and $\widehat{\mathcal{E}}_6$ ($\Hiss_6$). 
The quantity $\widehat{\mathcal{E}}_{2n}^\bullet$ is an exact integral of motion for the model $\simpleH{2n}$ 
and its PDF has a null IQR.} 
\end{figure*}

To retrieve the long-term statistical behavior of the QIs, we consider the ensembles of 1080 numerical integrations 
of the dynamical models $\Hiss_4$ and $\Hiss_6$, with very close initial conditions and spanning 100 Gyr in the future, 
that have been presented in Ref.~\citep{Hoang2022}. 
We also consider the similar ensembles of solutions for the simplified Hamiltonians $\Huv_4^\bullet$ and $\Huv_6^\bullet$ 
that we introduce in Sect.~\ref{sec:new_truncation}. 
We report in Fig.~\ref{fig:PDF_QIs} the time evolution of the ensemble PDFs of the low-pass filtered dimensionless QIs and 
dimensionless actions $\Chi_1,\Psi_3$ for the different models (the cutoff frequency of the time filter is set to 1 Myr$^{-1}$, 
as in Sec.~\ref{sec:numerical_checks}). More precisely, to highlight the growth of the statistical dispersion, we consider 
at each time the PDF of the signed deviation from the ensemble mean, so that all the plotted distributions have a null mean. 
At each time, the PDF estimation takes into account only the stable orbital solutions, that is, those solutions 
whose running maximum of Mercury eccentricity is smaller than 0.7 \citep{Hoang2022}. Figure~\ref{fig:PDF_QIs} shows that 
the QIs are indeed slow quantities when compared to the LL action variables. 
The growth of the QI dispersion is detailed in Fig.~\ref{fig:IQR_QIs}, where we report the time evolution of the interquartile 
range (IQR) of their distributions. After a transient phase lasting about 100 Myr and characterized by the exponential separation of close trajectories, 
the time growth of the IQR follows a power law typical of diffusion processes. Figures~\ref{fig:PDF_QIs} and \ref{fig:IQR_QIs} 
clearly show the slower diffusion of $\Cinc$ and $\mathcal{C}_2$ in the model $\simpleH{2n}$ when compared to $\Hiss_{2n}$. 
We recall that $\mathcal{E}_{2n}^\bullet$ is an exact integral of motion for the model $\simpleH{2n}$ (see Sect.~\ref{sec:new_truncation}) 
and its PDF has null dispersion. 

\clearpage
\bibliography{iss3} 

\end{document}